%% file: combined.tex
\documentclass[
aip,jcp,superscriptaddress,
reprint, floatfix
]{revtex4-1}

\usepackage[version=3]{mhchem} 

\usepackage{dcolumn}
\usepackage{bm}
\usepackage{amsfonts}
\usepackage{bbm}
\usepackage{todonotes}
\usepackage{tikz}
\usepackage{cool}
\usepackage[fleqn]{mathtools}
\usepackage{isomath}
\usepackage{lmodern}
\usepackage{multirow}
\usepackage[normalem]{ulem}
\usepackage{caption}
\usepackage{subcaption}
\usepackage{multibib}

\definecolor{darkgreen}{rgb}{0.0, 0.5, 0.0}

\begin{document}

\input{title.tex}

\begin{abstract}
\input{abstract.tex}
\end{abstract}

\maketitle

\input{text.tex}

\section*{Acknowledgement}
\input{acknowledgements.tex}



\bibliography{refs_dmw,refs}

\clearpage

\onecolumngrid

\begin{center}
{\Large \textbf{Supplementary Information}}
\end{center}

\input{text_SI.tex}

\end{document}

%% file: title.tex
\title{Hydrogen Bonding and Nuclear Quantum Effects in Clays}

\author{Pawan K. J. Kurapothula}

\author{Sam Shepherd}

\author{David M. Wilkins}
\affiliation{Atomistic Simulation Centre, School of Mathematics and Physics, Queen’s University Belfast, Belfast BT7 1NN, Northern Ireland, United Kingdom}
\email{d.wilkins@qub.ac.uk}

%% file: abstract.tex
Hydrogen bonds are of paramount importance in the chemistry of clays, mediating the interaction between the clay surface and water, and for some materials between separate layers.
It is well-established that the accuracy of a computational model for clays depends on the level of theory at which the electronic structure is treated.
However, for hydrogen-bonded systems the motion of light H nuclei on the electronic potential energy surface is often affected by quantum delocalization.
Using path integral molecular dynamics, we show that nuclear quantum effects lead to a relatively small change in the structure of clays, but one that is comparable to the variation incurred by treating the clay at different levels of electronic structure theory.
Accounting for quantum effects weakens the hydrogen bonds in clays, with H-bonds between different layers of the clay affected more than those within the same layer; this is ascribed to the fact that the confinement of an H atom inside a layer is independent of its participation in hydrogen-bonding.
More importantly, the weakening of hydrogen bonds by nuclear quantum effects causes changes in the vibrational spectra of these systems, significantly shifting the O--H stretching peaks and meaning that in order to fully understand these spectra by computational modelling, both electronic and nuclear quantum effects must be included.
We show that after reparametrization of the popular CLAYFF model for clays,  the O--H stretching region of their vibrational spectra better matches the experimental one, with no detriment to the model's agreement with other experimental properties.

%% file: text.tex
\section{Introduction}\label{sec:introduction}

Clay minerals play a key role in applications including wastewater treatment~\cite{Bhattacharyya2008,struijk2017novel}, catalysis~\cite{Johns1979,Vaccari1999} and the separation of oil and water~\cite{Xue2013}.
Although a variety of experimental techniques have been brought to bear on the problem of understanding the structure and dynamics of clay systems, including X-ray and neutron diffraction~\cite{Martins2014,Zhou2018}, infrared spectroscopy~\cite{Madejova2017} and contact angle measurements~\cite{Pan2020},
the picture provided by experiments is generally incomplete.
It is difficult to disentangle the contributions of different types of clay surface and microscopic motifs~\cite{Harvey2019,Pan2020}, and the small size (sub-$\mu$m) and surface roughness of the grains of some clays means that standard experiments cannot resolve their structures~\cite{Yin2012}.
These factors make tasks such as pinpointing the positions of hydrogen atoms, which are effectively invisible in standard X-ray scattering experiments, particularly problematic~\cite{Bish1993,liang1998calculated}.

Theoretical calculations can be used to assist in interpreting experiments, giving access to microscopic information that is difficult or impossible to access otherwise;
in clays which may present different types of interface, these calculations are able to resolve the properties of individual interfaces.
A key example is kaolinite, whose sheets comprise (octahedral) aluminol and (tetrahedral) silica layers: while contact angle measurements indicate that it is strongly wetted by water~\cite{Pan2020}, atomistic simulations consistently show that the aluminol layers are hydrophilic, while the silica are hydrophobic~\cite{Tunega2002,Solc2011,Tian2017,chen2019mechanism}. This amphiphilicity cannot be straightforwardly detected by experiments.
The positions of H atoms, which are difficult to detect experimentally, can much more straightforwardly be inferred by combining theoretical and experimental methods~\cite{Giese1971,Collins1991,Collins1992,liang1998calculated,rocha2018revisiting}.

In turn, knowledge of the hydrogen atoms' positions gives a better understanding of hydrogen-bonding in clays.
H-bonding plays a crucial role in their properties, either in their interactions with water~\cite{warne2000computer,Tambach2006,Hu2008,Szczerba2016,chen2019mechanism} or, in the case of so-called 1:1 clays whose layers comprise an H-bond donating and an H-bond accepting sheet, in the cohesive forces holding these layers together~\cite{Cygan2004,Tunega2012,liang1998calculated}.
Theoretical studies have been carried out at multiple scales, from quantum-mechanical calculations intended to find the ground-state structure of clays and characterize the types of H-bonding~\cite{Chatterjee1998,Tunega2012,Alvim2015,Peng2016,Zen2016,Richard2019} to molecular dynamics simulations, either using \textit{ab initio} descriptions of the electron structure to study clay wetting~\cite{Tunega2002,Hu2008} or acid-base chemistry~\cite{Boek2003,Tazi2012} or larger-scale studies with classical forcefields~\cite{Cygan2004,Cygan2021} to model diffraction~\cite{Hande2018} or vibrational spectroscopy~\cite{Greathouse2009} experiments.

Regardless of the manner in which the system is described, theoretical approaches to describing them are by nature imperfect.
Classical forcefields are relatively computationally inexpensive, and have been used to model clays for long simulation times and large system sizes~\cite{Tambach2004, Dashtian2017, TeichMcGoldrick2015}. However, they generally suffer from a lack of transferability and may need to be reparametrized to describe certain physical effects, e.g. vibrations~\cite{Greathouse2009}.
On the other hand, quantum-mechanical treatments are generally more transferable, but require compromises to be made, either in the level of theory used, in the size of system investigated or in the timescales of events to be studied. While methods like DFT are well-established for describing H-bonded systems, the results of these calculations depend sensitively on the level of theory used~\cite{Tunega2012,Gillan2016,Zen2016,Marsalek2017,CrastoDeLima2017,Cheng2019}. In particular, the general consensus is that dispersion effects must be included to fully capture the physics of clays~\cite{Tunega2012,Zen2016}.
Despite great improvements in computational treatments of clay materials in recent years, current methods are not able to describe all experiments equally well.

H-bonds are mediated by interactions between light nuclei, whose motion can be influenced by nuclear quantum effects (NQEs) such as zero-point energy and quantum dispersion.
It is well established that 
in order to accurately model the structure, thermodynamics and dynamics of H-bonded systems, 
the quantum-mechanical nature not only of the electrons but also of the nuclear motion must be accounted for~\cite{Walker2010,Ceriotti2016,Gillan2016,Cahlik2021}.
Nuclear quantum effects have been shown to affect proton disordering in high-pressure portlandite~\cite{Dupuis2017a}, and to some extent NQEs have been accounted for in clays by applying zero-point corrections to the calculated ground-state energy~\cite{Criscenti2005,Kubicki2019}.
NQEs at finite temperatures have not hitherto been considered in the simulation of clays, but their inclusion represents a plausible candidate to improve the agreement between these simulations and the experimental results.

In this paper we gauge the effect of nuclear quantum fluctuations on the properties of two clay minerals at room temperature by applying classical and path integral molecular dynamics simulations.
Although a full treatment of these systems requires the use of electronic structure theory, \textit{ab initio} molecular dynamics (AIMD) simulations are computationally very expensive, and instead we use a classical force field, focussing instead only on the effect of including NQEs in isolation.
We show that the effect of quantum fluctuations on the structures of these clays is comparable to the differences incurred by using different DFT functionals, implying that accounting for the quantum-mechanical nature of nuclei is as important as correctly treating the electronic structure.
We note that the O--H stretching frequencies observed in path integral simulations are red-shifted relative to those of classical simulations, indicating that quantum vibrational spectra are significantly different from their classical counterparts.
Taken together, these results indicate that NQEs must be included for an accurate simulation of clay properties, and point towards the requirements for a suitable theoretical description.
Finally, we show that a reparametrization of the popular CLAYFF forcefield~\cite{Cygan2004,Greathouse2009,Pouvreau2019,Cygan2021} greatly improves the description of the O--H stretching region of the vibrational spectrum in calculations including NQEs.

\section{Simulation Details}\label{sec:methods}

Clay minerals have a layered structure, with each layer comprised of octahedral sheets (containing divalent or trivalent metals such as aluminium, iron or magnesium) and tetrahedral sheets (containing silicon), with oxygen atoms bridging the sheets~\cite{Greenwell2006}. The octahedral sheets also contain hydroxyl groups attached to the metal ions.
Perfect clays may be dioctahedral (with two thirds of octahedral sites occupied by trivalent metal ions) or trioctahedral (with all octahedral sites occupied by divalent metals).
Furthermore, the layers contain a single octahedral sheet attached either to one or two tetrahedral layers (described as 1:1 and 2:1 clays, respectively).
The forces holding together the layers depend on the type of clay: the octahedral sheets in 1:1 clays are terminated with O--H groups, which are able to form hydrogen bonds with the tetrahedral sheet of the neighbouring layer. On the other hand, 2:1 clays are terminated by silica groups, and neighbouring layers interact only via dispersion forces.

We study the effects of NQEs on two dioctahedral clay minerals: the 1:1 type clay kaolinite (Al$_{2}$Si$_{2}$O$_{5}$(OH)$_{4}$) and the 2:1 type clay pyrophyllite (Al$_{2}$Si$_{4}$O$_{10}$(OH)$_{2}$).
The comparison between 2:1 and 1:1 clays allows the effect of quantum fluctuations on inter-layer and intra-layer H-bonding to be isolated and studied separately, with only kaolinite exhibiting inter-layer hydrogen bonds.
It is well-established that the structure of clay minerals is highly sensitive to the level of theory used to describe the electronic structure~\cite{Tunega2012}, with the method most accurate for treating clays not yet determined.
The calculations which we carry out require long simulation times (on the order of hundreds of picoseconds), which makes the use of electronic structure theory prohibitively expensive.
Here, we focus only on the question of how important NQEs are in modelling clay structure and dynamics and use the CLAYFF forcefield~\cite{Cygan2004,Greathouse2009,Pouvreau2019,Cygan2021}.
The parameters of Refs.~\onlinecite{Cygan2004,Greathouse2009} were used, with the Morse potential for the O--H stretch~\cite{Greathouse2009} truncated at fourth order as in Ref.~\onlinecite{Habershon2009} to avoid the breaking of these bonds into unphysical products.
\begin{figure}[h!]
\centering
\includegraphics[width=\linewidth]{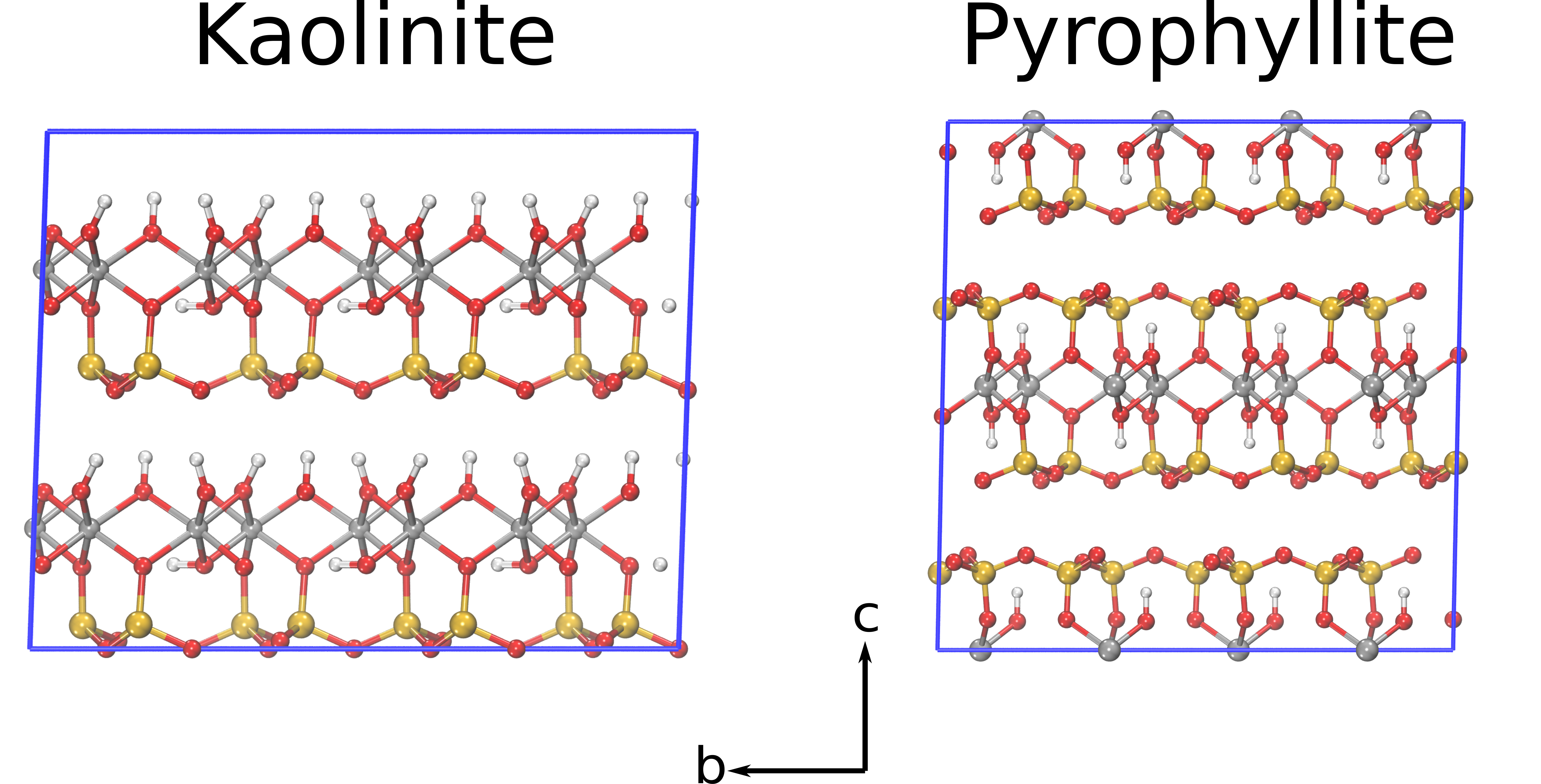}
\caption{
Experimental structures for the two clays studied in this work, taken from Refs.~\onlinecite{Bish1993} (kaolinite) and \onlinecite{Lee1981} (pyrophyllite). In all cases, the unit cell has been replicated twice in every direction to provide the starting point for simulations.
\label{fig:experimental_unit_cells}
}
\end{figure}

The starting structures for our simulations were taken from experimental data: that of pyrophyllite~\cite{Lee1981} was taken from X-ray diffraction results, and the structure of kaolinite~\cite{Bish1993} from neutron diffraction.
Fig.~\ref{fig:experimental_unit_cells} shows the experimental unit cells for these systems.
In all cases the unit cell was repeated twice in the direction of each unit cell vector to form a $2\times 2\times 2$ supercell.
Classical simulations began with a 50 ps equilibration in the NVT ensemble at 300 K, followed by a further equilibration for 100 ps in the constant-stress (NST) ensemble at 300 K, with diagonal elements of the stress tensor equal to 1 bar and off-diagonal elements equal to zero.
Static and structural properties were collected over a subsequent 500 ps NST production run.
In all cases, a timestep of 0.25 fs was used, the temperature was controlled by a fast-forward Langevin thermostat~\cite{Hijazi2018} and the stress controlled by the algorithm of Ref.~\onlinecite{Raiteri2011}.
Ten independent configurations were taken from the production calculations, their momenta resampled, and used as starting points for microcanonical (NVE) calculations of 100 ps, which were used to compute all reported dynamical properties.
\begin{table*}
\centering
\caption{
Unit cell parameters for kaolinite and pyrophyllite, as obtained from classical MD (CL) and PIMD (QM) calculations in the constant-stress (NST) ensemble.
Note that since a $2\times 2\times 2$ supercell is used, all lengths are divided by 2.
$\Delta_{\text{\tiny{QC}}}$ is the difference between quantum and classical calculations, $\sigma_{\text{\tiny{DFT}}}$ is the standard error in cell parameters from dispersion-corrected DFT functionals, computed using the results of Refs.~\onlinecite{Tunega2012} and \onlinecite{CrastoDeLima2017} (see SI for further details). The experimental value is given by EXP, with the results for kaolinite obtained as an average of the results from Refs.~\onlinecite{Neder1999,Bish1989} at 300 K, and the result for pyrophyllite obtained from~\onlinecite{Lee1981}.
In all cases, the uncertainty is in the fourth (for unit cell lengths) or third (for angles) decimal digit.
\label{tab:kaolinite_pyrophyllite}}
\begin{tabular}{c|ccccc|ccccc}
\hline\hline
& \multicolumn{5}{c|}{Kaolinite} & \multicolumn{5}{c}{Pyrophyllite} \\
Property & CL & QM & $\Delta_{\text{\tiny{QC}}}$ & $\sigma_{\text{\tiny{DFT}}}$ & EXP & CL & QM & $\Delta_{\text{\tiny{QC}}}$ & $\sigma_{\text{\tiny{DFT}}}$ & EXP \\
\hline
$a (\text{\AA})$ & 5.195 & 5.204 & 0.009 & 0.015 & 5.155 & 5.191 & 5.202 & 0.011 & 0.014 & 5.160 \\
$b (\text{\AA})$ & 8.950 & 8.969 & 0.019 & 0.024 & 8.944 & 9.013 & 9.033 & 0.020 & 0.036 & 8.966 \\
$c (\text{\AA})$ & 7.426 & 7.444 & 0.018 & 0.051 & 7.403 & 9.412 & 9.430 & 0.018 & 0.097 & 9.347 \\
$\alpha (^{\circ})$ & 91.68 & 91.71 & 0.03 & 0.13 & 91.70 & 91.44 & 91.44 & 0.00 & 0.05 & 91.18 \\
$\beta (^{\circ})$ &  104.83 & 104.82 & -0.01 & 0.15 & 104.74 & 98.82 & 98.96 & 0.04 & 0.07 & 100.46 \\
$\gamma (^{\circ})$ & 90.59 & 90.56 & -0.03 & 0.04 & 89.92 & 89.82 & 89.82 & 0.00 & 0.05 & 89.64 \\
\hline\hline
\end{tabular}
\end{table*}

NQEs on all static properties were accounted for using path integral molecular dynamics (PIMD)~\cite{Parrinello1984} with 32 ring-polymer replicas.
The simulation details for PIMD calculations were the same as for the classical case, but with the temperature controlled using the global path integral Langevin equation (PILE) thermostat~\cite{Ceriotti2010}.
Dynamical properties were carried out using thermostatted ring polymer molecular dynamics (TRPMD)~\cite{Habershon2013,Rossi2014}, with details the same as for the classical NVE runs.
All simulations were carried out using the i-PI wrapper~\cite{Ceriotti2013,Kapil2019} interfaced with LAMMPS~\cite{LAMMPS}.
Input files and starting configurations are provided in the supplementary information (SI).

\section{Results and Discussion}\label{sec:results}

\subsection{Clay Structure}

Table~\ref{tab:kaolinite_pyrophyllite} shows the unit cell parameters for kaolinite and pyrophyllite from classical and path-integral NST simulations.
Nuclear quantum effects cause the simulation box to increase in all directions, indicating that intralayer (parallel to the $a$ and $b$ directions) and interlayer (parallel to the $c$ direction) quantum effects are just as important in determining the overall effect.
The effect of NQEs on the structure of clays is very subtle, though similar to that on other hydrogen-bonded systems: quantum effects lead to a decrease in density of $\sim 0.6\%$ in clays, and $\sim 1\%$ in water~\cite{Cheng2019}.

The difference between quantum and classical structures should be compared to the spread of values obtained at different levels of electronic structure theory and to the deviation between the calculated and experimental structures.
Table~\ref{tab:kaolinite_pyrophyllite} compares $\Delta_{\text{\tiny{QC}}}$, the difference between quantum and classical values of a structural parameter, with $\sigma_{\text{\tiny{DFT}}}$, the standard error in this parameter among density functional theory calculations with dispersion corrections in Refs.~\onlinecite{Tunega2012,CrastoDeLima2017} (see SI for details). The difference between quantum and classical values is on the same order of magnitude as the spread of quantum-chemical values, indicating that a correct treatment of nuclear quantization is as important as a correct treatment of the electronic structure.
For all quantities, $\Delta_{\text{\tiny{QC}}} < \sigma_{\text{\tiny{DFT}}}$; this is particularly true for the unit cell $c$ length, normal to the layers, indicating that in order to correctly describe interlayer bonding it is more important to correctly account for the electronic structure than the nuclear quantization.

Ref.~\onlinecite{Tunega2012} showed that DFT with generalized gradient approximation (GGA) functionals predict unit cells that are larger than the experimental one for a variety of clays, with dispersion corrections improving the description to some degree but generally still leading to an under-binding.
The results of this section suggest that a DFT functional for accurately describing a clay in MD simulation should instead over-bind, giving a unit cell that is smaller than in experiment, which will be compensated for by thermal and nuclear quantum effects.

\begin{figure*}
\centering
    \begin{subfigure}[b]{0.49\textwidth}
        \centering 
        \includegraphics[scale=0.7]{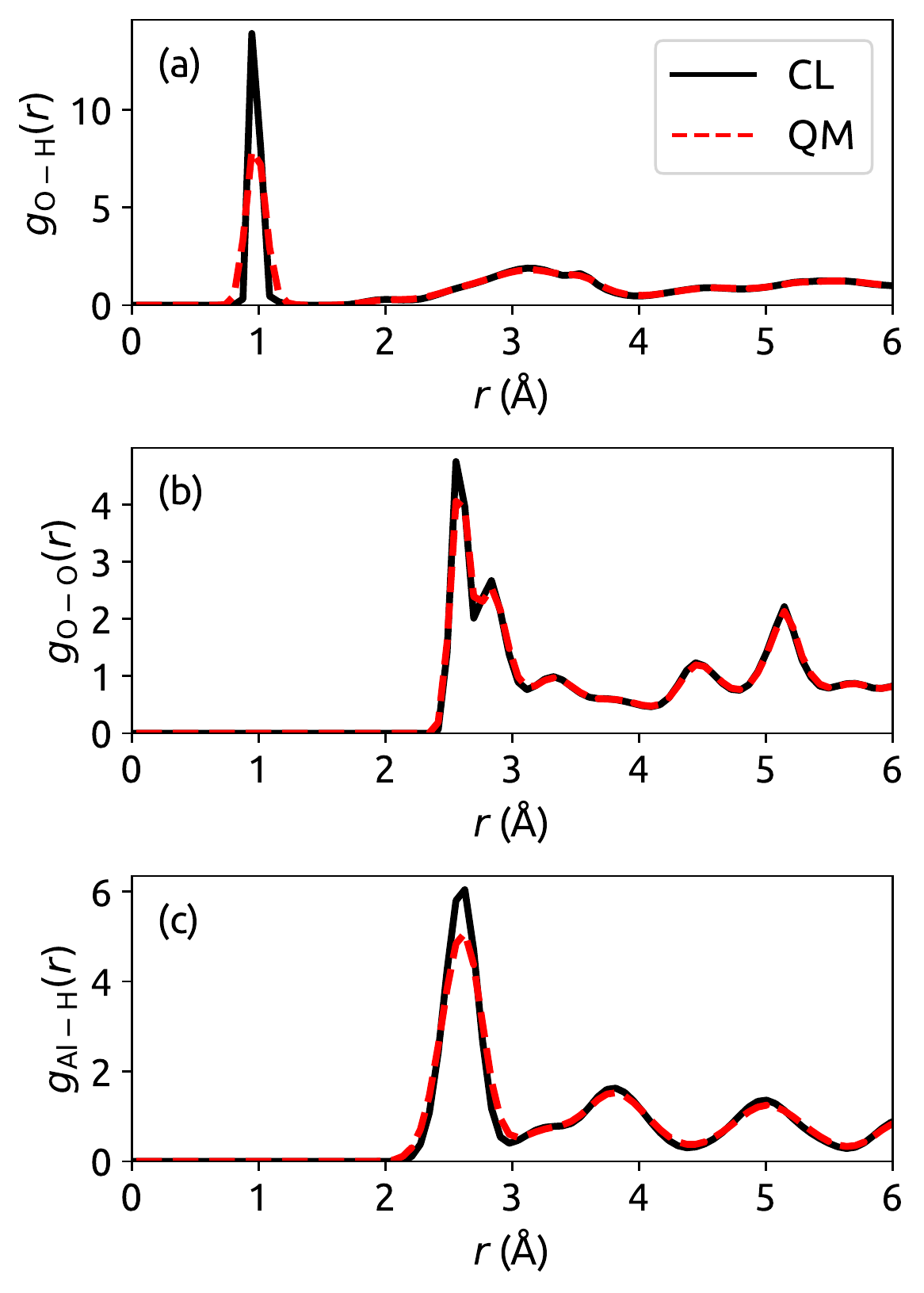}
        \caption{Kaolinite}
        \label{subfig:rdfs_kaolinite} 
    \end{subfigure}
    \begin{subfigure}[b]{0.49\textwidth}
        \centering 
        \includegraphics[scale=0.7]{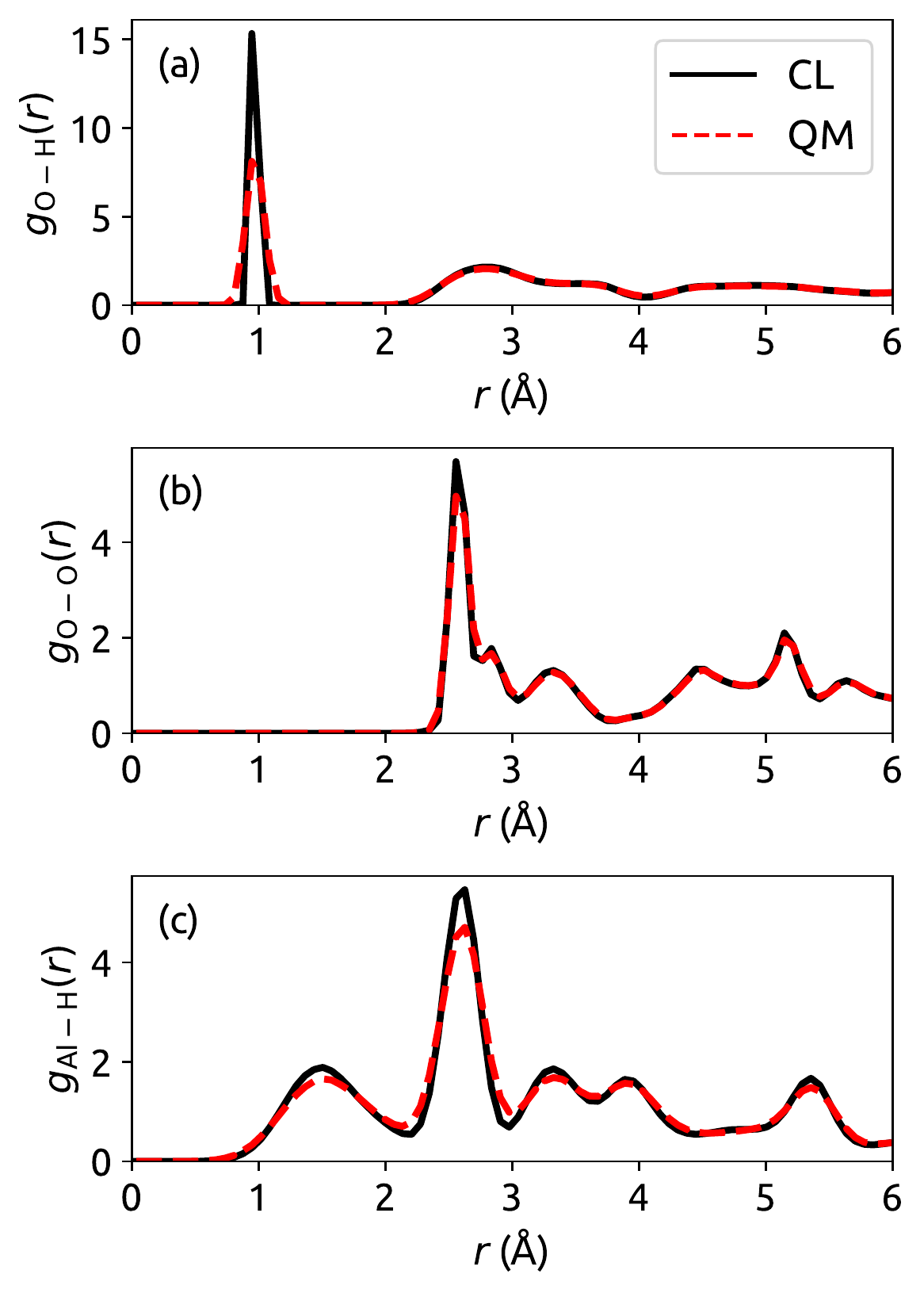}
        \caption{Pyrophyllite}
        \label{subfig:rdfs_pyrophyllite} 
    \end{subfigure}
    \caption{Radial distribution functions (RDFs) for (top) O--H, (middle) O--O and (bottom) Al--H pairs. Solid black lines show the RDF as calculated from classical MD and the dashed red lines show the RDF when including NQEs using PIMD calculations.} 
    \label{fig:rdfs_both_clays}
\end{figure*}

A more detailed picture of the structures of these two clays is given by their radial distribution functions (RDFs). 
Fig.~\ref{fig:rdfs_both_clays} shows the functions that are particularly affected by NQEs for kaolinite and pyrophyllite, with further RDFs shown in the SI.
Accounting for these effects leads to a de-structuring~\cite{Ceriotti2016}, with distribution functions becoming more spread.
There are noticeable effects not only for the O and H atoms, but also in $g_{\text{Al--H}}(r)$.
This reflects the fact that weaker H-bonds will lead to Al--O--H bond angles $\theta_{\text{\tiny{AlOH}}}$ that are less constrained by participation of the H atom in a hydrogen-bond, allowing for a slightly wider range of accessible Al--O--H bond angles: for classical kaolinite $\theta_{\text{\tiny{AlOH}}} = \left( 126 \pm 12\right)^{\circ}$ and for quantum kaolinite $\theta_{\text{\tiny{AlOH}}} = \left( 126 \pm 14\right)^{\circ}$, while for classical pyrophyllite $\theta_{\text{\tiny{AlOH}}} = \left( 123 \pm 12\right)^{\circ}$ and for quantum pyrophyllite $\theta_{\text{\tiny{AlOH}}} = \left( 123 \pm 13\right)^{\circ}$, indicating that the greater propensity of H-bonds in quantum-mechanical simulations to break in directions perpendicular to the O--H bond leads to a greater uncertainty in the geometry of the Al--O--H triplet, although the average geometry is unchanged by nuclear quantum fluctuations.

The results for the structural characterization of kaolinite and pyrophyllite suggest that structural features that are not related to H-bonding are not directly affected by NQEs. In order to better understand the effect of nuclear quantum fluctuations on clays, we focus now on hydrogen-bonding.
H-bonds were identified using the geometric criterion of Luzar and Chandler~\cite{Luzar1996a}; that is, two water molecules are hydrogen-bonded if their oxygen atoms are separated by less than 3.5~\AA and there is an O-O-H angle less than 30$^{\circ}$ for an H atom attached to one of the oxygens.

Table~\ref{tab:hbonds} shows the fraction of intact hydrogen bonds and dangling O--H bonds in the two clays from classical and path-integral calculations, with the results for kaolinite split into inter-layer and intra-layer H-bonds.
\begin{table}[t!]
\centering
\caption{
Number of hydrogen bonds that are intact for kaolinite and pyrophyllite, from classical MD (CL) and path integral MD (QM) calculations. For kaolinite, H-bonds are classified as being intralayer or interlayer. ``$T=0~\text{K}$ Intralayer'' H-bonds are intralayer hydrogen bonds that exist in the experimental structure at zero temperature, while ``$T=300~\text{K}$ Intralayer'' H-bonds are intralayer hydrogen bonds formed by O--H bonds that at zero temperature participate in \textit{interlayer} H-bonding.
For pyrophyllite, all H-bonds are intralayer.
\label{tab:hbonds}}
\begin{tabular}{c|cc}
\hline\hline
\multicolumn{3}{c}{Kaolinite} \\
\hline\hline
Type & CL & QM \\
Interlayer & 34.9 & 32.9 \\
$T=0~\text{K}$ Intralayer & 12.5 & 12.2 \\
$T=300~\text{K}$ Intralayer & 6.1 & 6.1 \\
\hline\hline
\multicolumn{3}{c}{Pyrophyllite} \\
\hline\hline
Type & CL & QM \\
Intralayer & 29.3 & 28.6 \\
\hline\hline
\end{tabular}
\end{table}
In the experimental structure of kaolinite, there are 48 O--H bonds participating in interlayer H-bonding and 16 bonds participating in intralayer H-bonding, with each of the latter donating a hydrogen-bond to two acceptors, for a total of 32 intralayer H-bonds. At room temperature, only $\sim 70\%$ of interlayer hydrogen-bonds and $\sim 60\%$ of intralayer hydrogen-bonds are intact, a marked difference from the ground state. The majority of intralayer O--H bonds now participate in only a single H-bond, indicating simply that thermal disordering has broken the symmetry between pairs of possible acceptors that existing in the ground state.
In addition, it can be seen that one-third of the intralayer hydrogen-bonds in kaolinite at room temperature are donated by O--H bonds that participate in interlayer H-bonding in the ground state, indicating that a significant number of H-bonds at the surface are able to switch between the two forms.
Pyrophyllite has 32 O--H bonds, most of which are participating in intralayer H-bonding at room temperature.
Table~\ref{tab:hbonds} indicates that nuclear quantum fluctuations affect the hydrogen-bonds holding together different layers significantly more than those between oxygen atoms in the same layer, with a $1\%$ and $2\%$ difference in numbers of intralayer H-bonds upon quantization of kaolinite and pyrophyllite respectively, and a $6\%$ difference for the interlayer H-bonds in kaolinite.

To understand why the major part of the quantum effect in these clays is in the interlayer interactions, we computed the centroid virial kinetic energy tensor,
\begin{equation}
\mathcal{T}_{ij} = \frac{1}{2}\delta_{ij} k_{\rm B} T - \frac{1}{4n}\sum_{k=1}^{n} \left[  \left(q_{k,i} - \bar{q}_{i}\right)f_{k,j} + \left( q_{k,j} - \bar{q}_{j}\right)f_{k,i} \vphantom{\frac{1}{2}}\right],
\end{equation}
for H atoms,
where $\delta_{ij}$ is the Kronecker delta function, $k_{\rm B}$ the Boltzmann constant and $T$ the temperature; $q_{k,i}$ is the $i^{\rm th}$ Cartesian component of the position of the $k^{\rm th}$ ring-polymer replica of the atom, $\bar{q}_{i} = \frac{1}{n}\sum_{k=1}^{n} q_{k,i}$ the centroid of this atom and $f_{k,i}$ the $i^{\rm th}$ Cartesian component of the force on the $k^{\rm th}$ replica. $n$ is the number of replicas in the PIMD simulation.
The higher the kinetic energy of an atom, the more confined it is and the larger the zero-point energy contribution~\cite{Markland2012,Liu2013,Wilkins2015}. Calculation of the tensor $\boldsymbol{\mathcal{T}}$ allows the contribution of motion in different directions to be resolved.
\begin{table}[htb!]
\centering
\caption{
Components of the centroid virial kinetic energy tensor $\boldsymbol{\mathcal{T}}$ for H atoms in intact hydrogen bonds or part of a dangling O--H bond, for intralayer and interlayer hydrogen-bonding in kaolinite and for intralayer hydrogen-bonding in pyrophyllite.
$\mathcal{T}_{\parallel}$ is the component parallel to the O--H bond, $\mathcal{T}_{\perp}$ the sum of the two components perpendicular to the bond, and $\mathcal{T}=\mathcal{T}_{\parallel} + \mathcal{T}_{\perp}$.
All values are in meV.
The bottom panel shows the kinetic energy tensor for an H atom participating in an interlayer H-bond in kaolinite, with the component parallel to the O--H bond and one perpendicular component highlighted.
$\Delta \mathcal{T}_{i}$ is the change in a component of the kinetic energy tensor when an H-bond is broken.
\label{tab:kinetic_tensor}}
\begin{tabular}{c|ccc|ccc|ccc}
\hline\hline
\multicolumn{10}{c}{Kaolinite} \\
\hline\hline
& \multicolumn{3}{c|}{H-Bond} & \multicolumn{3}{c|}{Dangling Bond} \\
\hline
Type & $T_{\parallel}$ & $\mathcal{T}_{\perp}$ & $\mathcal{T}$ & $\mathcal{T}_{\parallel}$ & $\mathcal{T}_{\perp}$ & $\mathcal{T}$ & $\Delta\mathcal{T}_{\parallel}$ & $\Delta\mathcal{T}_{\perp}$ & $\Delta\mathcal{T}$\\
Intralayer & 100.7 & 40.6 & 141.3 & 100.4 & 40.9 & 141.3 & -0.3 & 0.3 & 0.0\\
Interlayer & 100.7 & 40.3 & 141.0 & 102.2 & 35.8 & 138.0 & 1.4 & -4.5 & -3.0\\
\hline\hline
\multicolumn{10}{c}{Pyrophyllite} \\
\hline\hline
& \multicolumn{3}{c|}{H-Bond} & \multicolumn{3}{c|}{Dangling Bond} \\
\hline
Type & $\mathcal{T}_{\parallel}$ & $\mathcal{T}_{\perp}$ & $\mathcal{T}$ & $\mathcal{T}_{\parallel}$ & $\mathcal{T}_{\perp}$ & $\mathcal{T}$ & $\Delta\mathcal{T}_{\parallel}$ & $\Delta\mathcal{T}_{\perp}$ & $\Delta\mathcal{T}$\\
Intralayer & 101.9 & 37.7 & 139.6 & 102.1 & 37.1 & 139.2 & 0.2 & -0.6 & -0.4\\
\hline\hline
\end{tabular}
\includegraphics[width=0.9\linewidth]{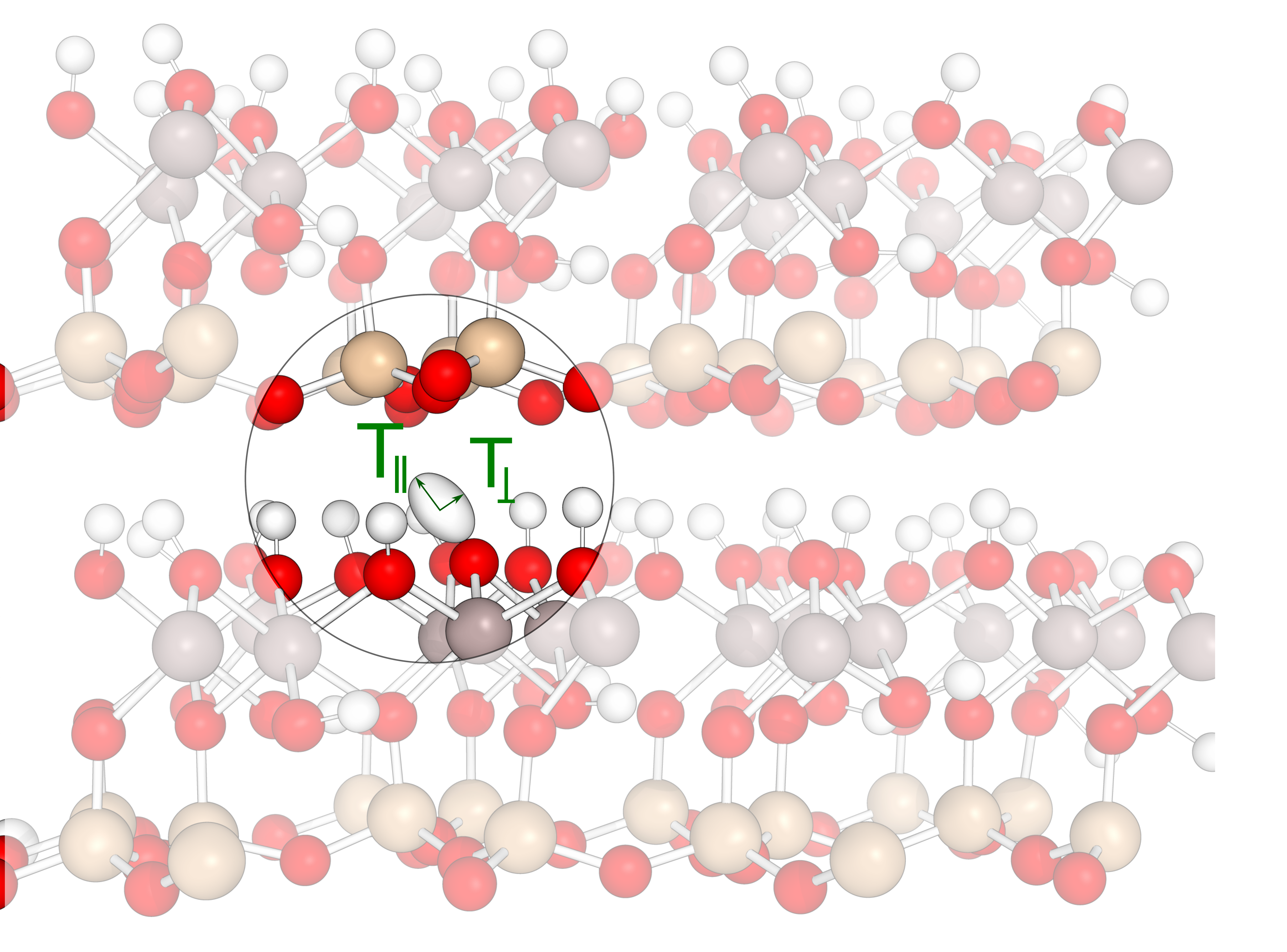}
\end{table}

Table~\ref{tab:kinetic_tensor} shows the components of the kinetic energy tensor of H atoms either involved in hydrogen bonds or not involved in hydrogen bonding, termed ``dangling'' O--H bonds (in both cases, split by whether they are intralayer or interlayer H-bonds or dangling bonds).
For each type of H atom three quantities are recorded: $\mathcal{T} = \text{tr}\left[\boldsymbol{\mathcal{T}}\right]$, the total kinetic energy (trace of the centroid virial tensor), $\mathcal{T}_{\parallel} = \hat{\boldsymbol{r}}_{\text{OH}}\cdot\boldsymbol{\mathcal{T}}\cdot\hat{\boldsymbol{r}}_{\text{OH}}$, the component parallel to the O--H bond and $\mathcal{T}_{\perp} = \mathcal{T} - \mathcal{T}_{\parallel}$, the component perpendicular to the bond.
In all cases, the breaking of a hydrogen bond decreases the kinetic energy component parallel to the O--H bond and increases the component perpendicular to the bond, reflecting the fact that in a dangling O--H bond the uncertainty of the H atom's position in the direction of the bond is decreased when it is participating in an H-bond and it is more confined in this direction; on the other hand, the bond itself is more free to rotate and the confinement in directions orthogonal to the bond is less~\cite{Wilkins2015}.
For intralayer H-bonds, these changes individually are relatively small ($\sim 0.4~\text{meV}$) and the competition of the two effects leads to a very small or negligible (in the case of kaolinite) change in the total kinetic energy.
This is because, even when the H-bonds are broken, the H atom is still in a confined environment, lowering the incentive for these bonds to break.
On the other hand, for interlayer H-bonds both the changes in individual components and in the total kinetic energy are an order of magnitude larger, indicating that H atoms in dangling interlayer O--H bonds are significantly less confined than those participating in interlayer H-bonds, adding an extra driving force for their breaking.
For this reason, the majority of the quantum effects in these clays are in the \textit{interlayer} hydrogen-bonds.
This analysis does not give an estimate of the total energy change upon breaking an H-bond, which is also affected by potential energy, but rather of the effect of changes in the confinement of the H atom on this energy.

\subsection{H-Bond Dynamics}

\begin{figure}[ht!]
\centering
\includegraphics[width=\linewidth]{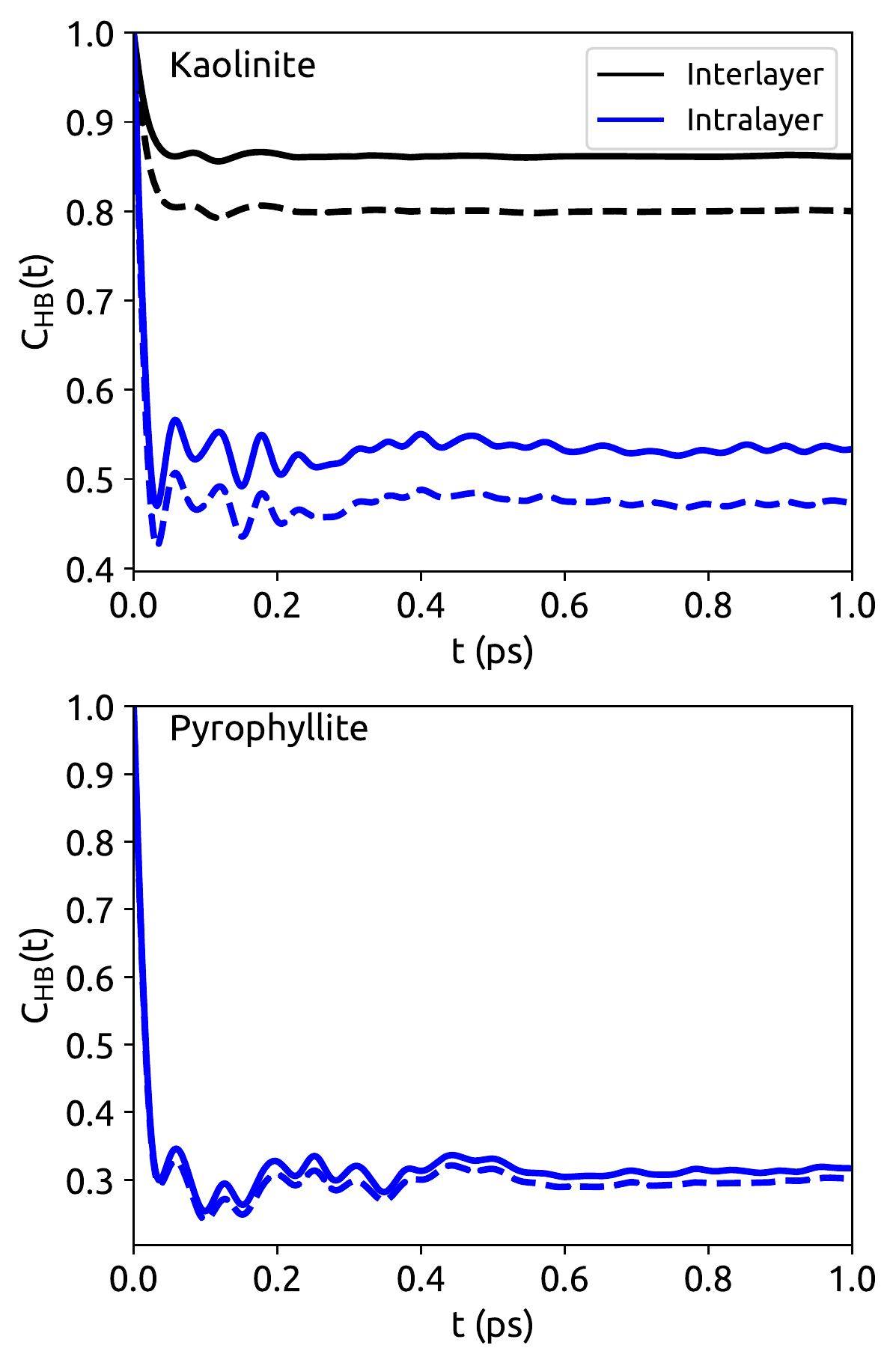}
\caption{
H-bond population correlation function of Eq.~\eqref{eq:hbond_function}, for kaolinite (top panel) and pyrophyllite (bottom panel). Black lines give this function for interlayer hydrogen bonds, where present, and blue lines for intralayer hydrogen bonds.
Solid lines show the results of classical MD calculations and dashed lines the results of TRPMD.
\label{fig:hbond_correlation_function}
}
\end{figure}

The H-bond population correlation function~\cite{Luzar1996a},
\begin{equation}\label{eq:hbond_function}
C_{\text{\tiny{HB}}}(t) = \frac{\left\langle h(0) h(t)\right\rangle}{\left\langle h(0)^{2}\right\rangle},
\end{equation}
where the average is over all possible H-bonding pairs, with $h(t) = 1$ if the pair is hydrogen-bonding and $h(t) = 0$ otherwise, describes the time-dependent probability that an initially intact H-bond will remain so.
For TRPMD calculations, $h(t)$ is calculated using the centroids of ring polymers.
Fig.~\ref{fig:hbond_correlation_function} shows $C_{\text{\tiny{HB}}}(t)$ for the two clays, with the hydrogen-bonds in kaolinite divided up into intra- and interlayer H-bonds.
The correlation function decays to a nonzero value, indicating that hydrogen bonds that break may re-form;
NQEs decrease this plateau value for kaolinite significantly more than for pyrophyllite, even in the case of intralayer H-bonds. This can be attributed to the fact that some O--H bonds in kaolinite are able to switch between interlayer and intralayer H-bonding, meaning that those which start off participating in an intralayer H-bond may later be participating in interlayer H-bonding -- and in doing so, incurring a greater quantum effect -- during the decay of the correlation function.
In the SI, we show that when the correlation function is calculated only for O--H bonds that only participate in intralayer H-bonding, the results are much closer to those of pyrophyllite.

The plateau value $C_{\text{\tiny{HB}}}(t\rightarrow\infty)$ for H-bonds follows the order $\text{pyrophyllite} < \text{kaolinite intralayer} < \text{kaolinite interlayer}$.
The large value for interlayer H-bonds in kaolinite shows that a large proportion of hydrogen bonds re-form after breaking, which can be attributed to the fact that layers in kaolinite were not observed to slide over each other on the simulation timescales used. This leaves interlayer H-bonds unable to reach any acceptor groups other than the initial one.
On the other hand, intralayer H-bonding generally takes place in cavities within which it is possible for a hydrogen-bond donor to change from one O acceptor to another, meaning that an intralayer H-bond is less likely to reform with the same initial receptor.

\subsection{Vibrational Spectra}

The picture of clays painted by the result thus far indicates that although an accurate modelling of their structure will require NQEs to be accounted for, their effects are fairly subtle. However, relatively small effects on the structure of a hydrogen-bonded system can translate into large effects in its vibrational density of states and thus in the resulting vibrational spectrum~\cite{Marsalek2017}.
The CLAYFF forcefield used in this work has been shown previously to predict O--H peaks whose frequencies are too low~\cite{Greathouse2009}, entirely consistent with the fact that the forcefield underestimates the strength of H-bonds.
While this is a relatively minor problem for the pure clay, in which the peaks can be ascribed straightforwardly to microscopic features and their absolute position does not matter, the addition of a solvent to the system would make it critical to obtain the correct spectrum for the clay part of the system.
We therefore move on to gauge the effect of nuclear quantization on the vibrational spectra of pure clays.

\begin{figure*}[ht!]
\centering
\includegraphics[width=\linewidth]{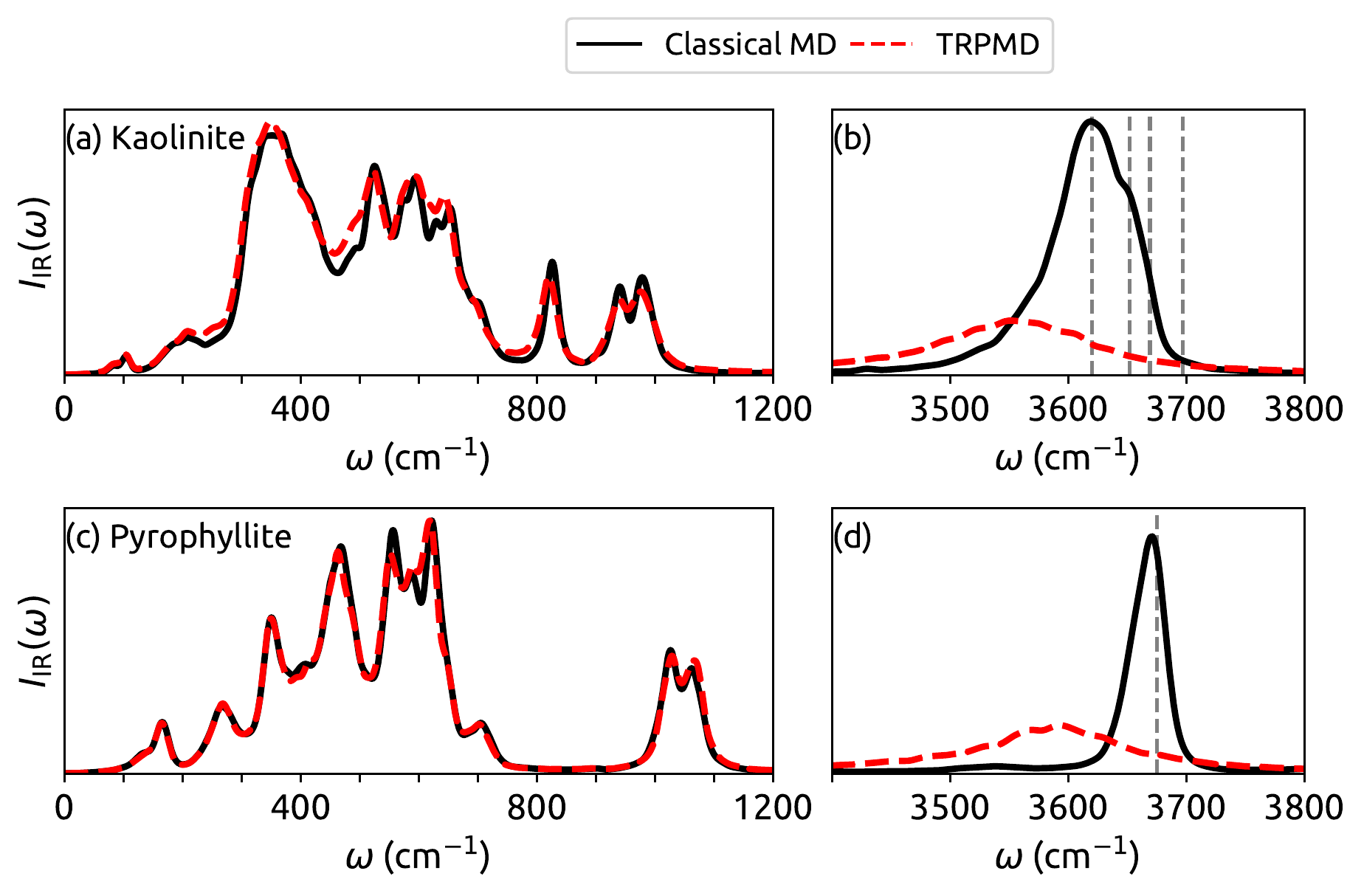}
\caption{
Infrared spectrum of kaolinite (top panels) and pyrophyllite (bottom panels). (a) and (c) show the low-frequency region and (b) and (d) the high-frequency (O--H stretching) region of the spectrum.
Solid black lines show the VDOS from classical MD simulations and dashed red lines the VDOS from TRPMD calculations. Dashed vertical lines give the experimental results of Ref.~\onlinecite{Farmer1974}.
In all cases, the spectra have been renormalized to have unit area.
\label{fig:IR_spectra}
}
\end{figure*}
We computed the infrared (IR) spectrum~\cite{McQuarrie2000},
\begin{equation}
I_{\text{IR}}(\omega) \sim \omega^{2} \int \left\langle \boldsymbol{P}(0)\cdot\boldsymbol{P}(t)\right\rangle e^{i\omega t}\,{\rm d} t,
\end{equation}
where $\boldsymbol{P}(t)$ is the polarization (macroscopic dipole moment) at time $t$, given by $\boldsymbol{P}(t) = \frac{1}{V}\sum_{i} q_{i} \boldsymbol{r}_{i}(t)$, with $V$ the system's volume, $q_{i}$ the partial charge of the $i^{\rm th}$ atom from the CLAYFF model and $\boldsymbol{r}_{i}(t)$ its position at time $t$. For TRPMD calculations, the centroid of the ring polymer is used to calculate the polarization, as in Ref.~\cite{Kapil2020}. In all cases, the spectrum is normalized to unit integral.

Fig.~\ref{fig:IR_spectra} gives the IR spectrum for kaolinite and pyrophyllite using classical MD and TRPMD. While the low-frequency modes are mainly unaffected, aside from a small change in intensity for kaolinite, nuclear quantum effects weaken hydrogen bonds, causing the O--H stretching peaks to be red-shifted.
For pyrophyllite, whose in-phase O--H stretching mode was used to parametrize the Morse potential of Ref.~\onlinecite{Greathouse2009}, this shift moves the peak away from the experimental value. For kaolinite, the classical peaks frequency, which is already somewhat underestimated in the classical simulations, is shifted further away from experiments when NQEs are included.
The O--H stretching peak is also broadened in the quantum dynamical calculations, which we attribute to a combination of quantum-mechanical dispersion and the known (unphysical) broadening of spectral peaks inherent to the TRPMD method.

\subsection{Modified Forcefield}

The results of the previous section show that NQEs only significantly affect the O--H stretching region of the vibrational spectra of clays.
For the CLAYFF forcefield, classical simulations give a much better agreement with experiments than those incorporating quantum effects; however, this is because the O--H bond terms in this forcefield were parametrized using classical simulations~\cite{Greathouse2009}.
Given that NQEs must be accounted for in order to accurately capture the properties of water, their inclusion is also needed to capture the properties of aqueous interfaces with clays, and any simulation of the water-clay surface would ideally account for nuclear quantization.

While a treatment of clays based on \textit{ab initio} MD combined with path integral methods would be preferable, this is often too computationally expensive to be feasible, particularly when studying vibrational spectra. However,
our results point to a simple improvement to the CLAYFF forcefield which will make it compatible with path integral simulations, without losing out on the inherently good agreement that it has with experimental results: that is, to increase the O--H stretching frequency in the model.
As pointed out in Ref.~\cite{Habershon2009}, since the forcefield is parametrized using experiments, in which nuclear quantum effects are present, it is optimal to use simulations in which NQEs are accounted for.
The relatively small difference between the results of classical and quantum simulations for other properties, in which the main change is in the strength of H-bonds, suggests that modifying this strength via the O--H stretch will not appreciably affect these properties.

\begin{figure}[ht!]
\centering
\includegraphics[width=\linewidth]{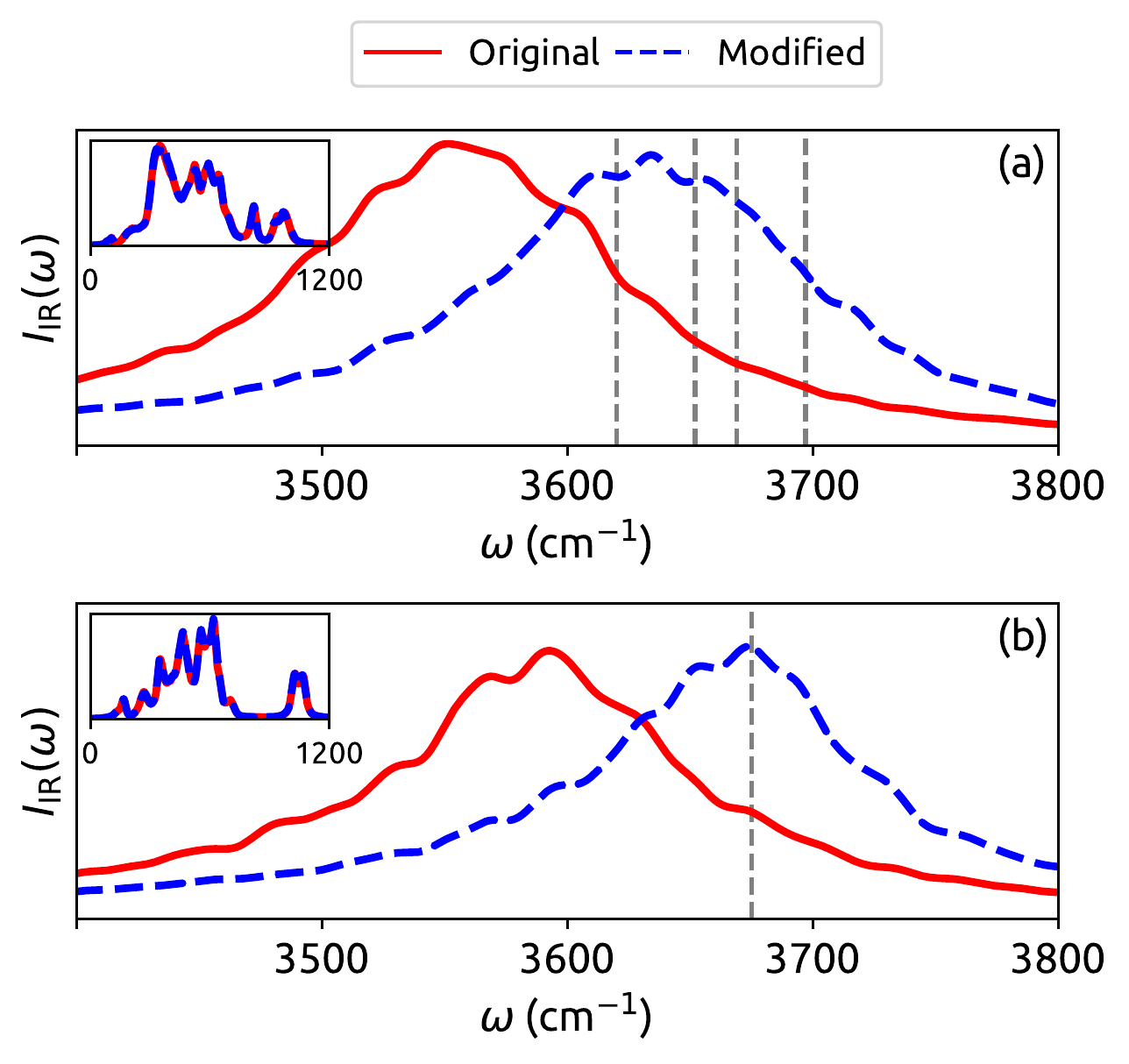}
\caption{
High-frequency region of the infrared spectrum of (a) kaolinite and (b) pyrophyllite. Insets show the density of states in the low-frequency region. Solid red lines show the results obtained using the original ClayFF model, and dashed blue lines show the results from a modified force field in which the O--H force constant is increased by 2\%. All results were obtained from TRPMD calculations.
Dashed vertical lines give the experimental results of Ref.~\onlinecite{Farmer1974}.
In all cases, the spectra have been renormalized to have unit area.
\label{fig:modified_FF_spectra}
}
\end{figure}

In Fig.~\ref{fig:modified_FF_spectra} we show that increasing the force constant for the O--H bond stretch by 2\% does indeed result in a much better agreement with the experimental vibrational spectra in the stretching region, leaving the remainder of the spectrum unchanged.
In the SI, we show that this change makes little difference to the structure of the two clays. 
This modification of the forcefield has essentially no effect on the quantum kinetic energy of H atoms in the directions perpendicular to H-bonds $\mathcal{T}_{\perp}$, and leads to an increase in the parallel component $\mathcal{T}_{\parallel}$. The difference $\Delta\mathcal{T}_{\parallel}$ when an interlayer hydrogen-bond is broken is the same in the original CLAYFF model and in this modified version, suggesting that there is little difference in overall H-bond strengths between the two models, and therefore that both models possess similar proclivities for H-bond breaking.
We refer to this modified forcefield as CLAYFF-TRPMD.

\section{Conclusions}\label{sec:conclusions}

In this paper, we have shown that nuclear quantum effects, while unnecessary for a quantitatively accurate description of the static properties of most pure clays, are key in describing their dynamics, particularly that of hydrogen-bonds and vibrational spectra, and thus in helping to interpret the experimental spectrum. This point will be even more important in more complex systems such as clay-water interfaces, in which several vibrational features overlap.
Future work will focus on investigating the bearing these conclusions have on the dynamics of intercalation between clay layers, as well as studying the spectra of aqueous clay systems with first-principles accuracy.

\section*{Supplementary Information}

The Supplementary Information (SI) contains input files and starting configurations for the simulations carried out in this work, as well as further information on static and dynamical quantities of the CLAYFF model with and without nuclear quantum effects.

%% file: acknowledgements.tex
D.M.W. acknowledges startup funding from Queen's University Belfast (QUB).
We are also grateful to the UK Materials and Molecular Modelling Hub for computational resorces, which is partially funded by EPSRC (EP/P020194/1 and
EP/T022213/1),
and for computer time from QUB's Research Computing Centre.
The authors thank Venkat Kapil, Yair Litman and Lorenzo Stella for critical reading of the manuscript.

%% file: text_SI.tex
\section*{Uncertainties in DFT Data}

Ref.~\onlinecite{Tunega2012} showed that when using density functional theory (DFT) to find the structure of clays, dispersion corrections are required for the best performance for a variety of materials (trioctahedral and dioctahedral, 1:1 and 2:1 clays). To find the uncertainty in the structural parameters, we took the unit cell lengths and angles from dispersion-corrected calculations in Refs.~\onlinecite{Tunega2012} and \onlinecite{CrastoDeLima2017}, as shown in Table~\ref{tab:uncertainties}.
\begin{table}[h!]
\caption{
Structural parameters for kaolinite and pyrophyllite obtained using density functional theory with the generalized gradient approximation and dispersion corrections. Results are either obtained from the work of Tunega \textit{et al.}~\cite{Tunega2012} or Crasto de Lima \textit{et al.}~\cite{CrastoDeLima2017}. Note that the precision of the reported results differs between references. $\sigma_{\text{\tiny{DFT}}}$ is the standard error among the different results.
\label{tab:uncertainties}}
\centering
\begin{tabular}{c||c|c|c|c|c|c}
\hline\hline
 & \multicolumn{6}{c}{Kaolinite} \\
\hline
Functional & PW91-D2\cite{Tunega2012} & PBE-D2\cite{Tunega2012} & vdW-TS\cite{Tunega2012} & RPBE-D2\cite{Tunega2012} & revPBE-vdW\cite{CrastoDeLima2017} & $\sigma_{\text{\tiny{DFT}}}$ \\
\hline
$a$ (\AA) & 5.170 & 5.177 & 5.176 & 5.198 & 5.25 & 0.015 \\
$b$ (\AA) & 8.971 & 8.983 & 8.971 & 9.019 & 9.10 & 0.024 \\
$c$ (\AA) & 7.316 & 7.313 & 7.329 & 7.384 & 7.58 & 0.051 \\
$\alpha~(\text{deg})$ & 92.36 & 91.76 & 91.78 & 92.30 & 92 & 0.13 \\
$\beta~(\text{deg})$  & 105.43 & 105.06 & 105.00 & 104.47 & 105 & 0.15 \\
$\gamma~(\text{deg})$ & 89.84 & 89.83 & 89.96 & 89.83 & 90 & 0.04 \\
\hline\hline
 & \multicolumn{6}{c}{Pyrophyllite}\\
\hline
Functional & PW91-D2\cite{Tunega2012} & PBE-D2\cite{Tunega2012} & vdW-TS\cite{Tunega2012} & RPBE-D2\cite{Tunega2012} & revPBE-vdW\cite{CrastoDeLima2017} & $\sigma_{\text{\tiny{DFT}}}$ \\
\hline
$a$ (\AA) & 5.168 & 5.167 & 5.174 & 5.180 & 5.24 & 0.014 \\
$b$ (\AA) & 8.979 & 8.978 & 8.972 & 9.003 & 9.16 & 0.036 \\
$c$ (\AA) & 9.302 & 9.300 & 9.360 & 9.364 & 9.81 & 0.097 \\
$\alpha~(\text{deg})$ & 90.97 & 90.97 & 90.76 & 90.90 & -- & 0.05 \\
$\beta~(\text{deg})$  & 100.90 & 100.91 & 100.70 & 100.66 & -- & 0.07 \\
$\gamma~(\text{deg})$ & 89.83 & 89.83 & 90.00 & 89.80 & -- & 0.05 \\
\hline\hline
\end{tabular}
\end{table}

\section*{Radial Distribution Functions}

\begin{figure*}
\centering
    \begin{subfigure}[b]{0.49\textwidth}
        \centering 
        \includegraphics[scale=0.7]{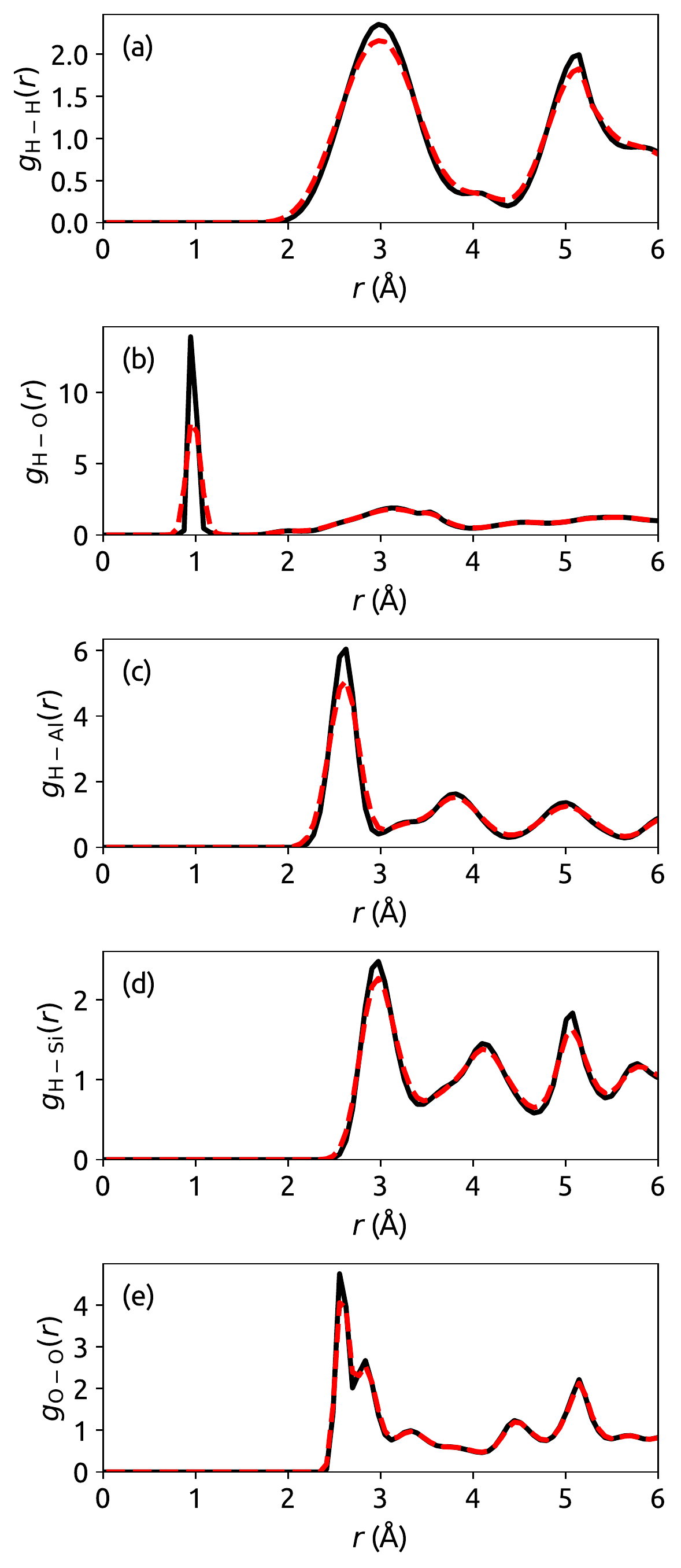}
        \label{subfig:rdfs_kaolinite_si_1} 
    \end{subfigure}
    \begin{subfigure}[b]{0.49\textwidth}
        \centering 
        \includegraphics[scale=0.7]{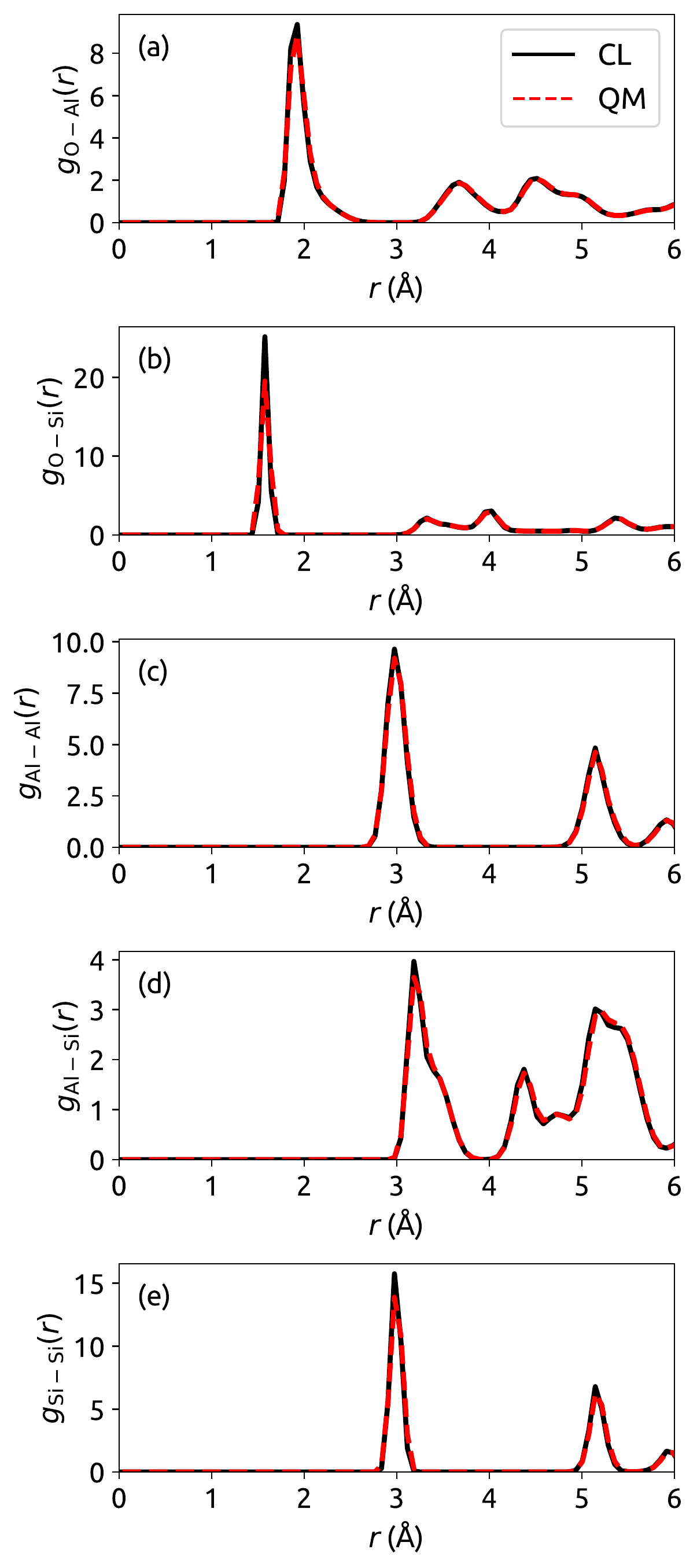}
        \label{subfig:rdfs_kaolinite_si_2} 
    \end{subfigure}
    \caption{
    Radial distribution functions (RDFs) for all element pairs in kaolinite. Solid black lines show the RDF as calculated from classical molecular dynamics and dashed red lines show the RDF when calculated from path integral molecular dynamics.
    } 
    \label{fig:rdfs_kaolinite_si}
\end{figure*}

\begin{figure*}
\centering
    \begin{subfigure}[b]{0.49\textwidth}
        \centering 
        \includegraphics[scale=0.7]{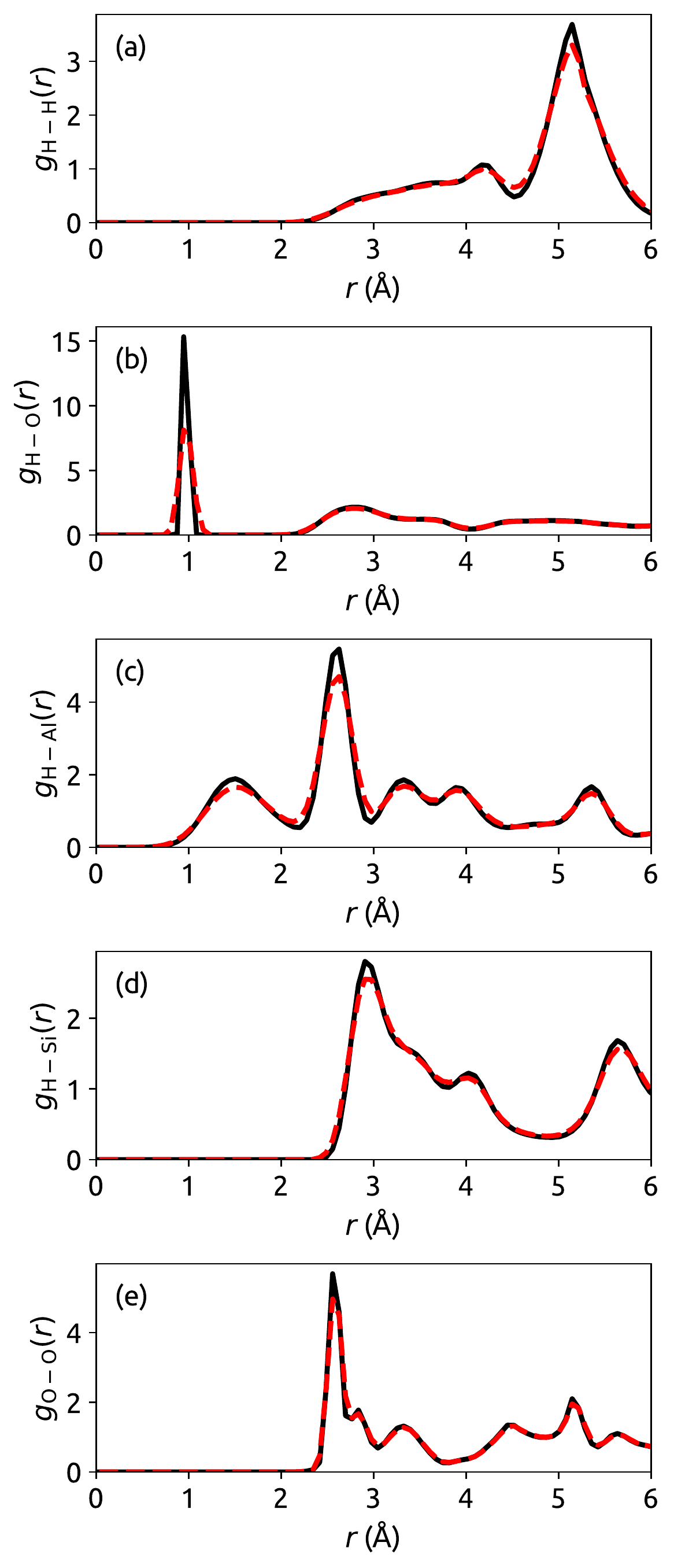}
        \label{subfig:rdfs_pyrophyllite_si_1} 
    \end{subfigure}
    \begin{subfigure}[b]{0.49\textwidth}
        \centering 
        \includegraphics[scale=0.7]{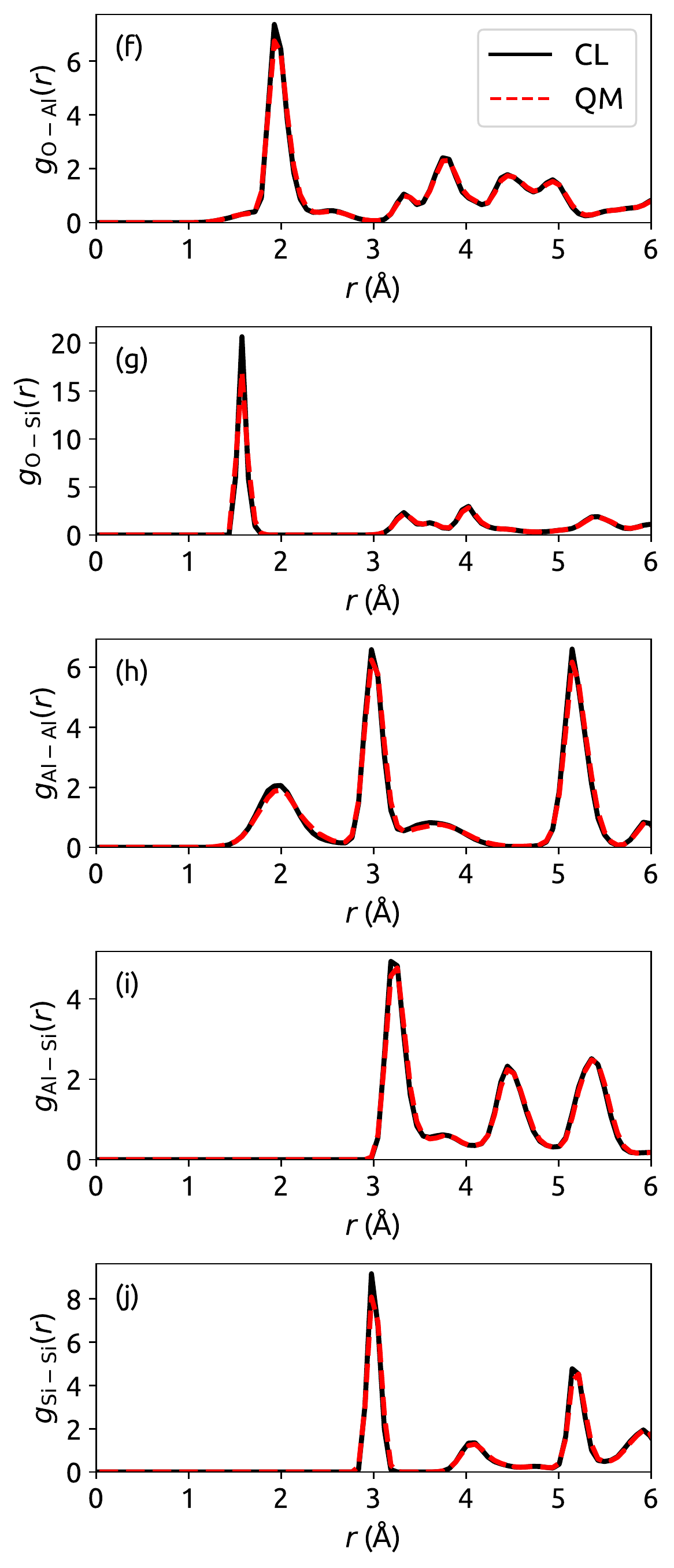}
        \label{subfig:rdfs_pyrophyllite_si_2} 
    \end{subfigure}
    \caption{
    Radial distribution functions (RDFs) for all element pairs in pyrophyllite. Solid black lines show the RDF as calculated from classical molecular dynamics and dashed red lines show the RDF when calculated from path integral molecular dynamics.
    } 
    \label{fig:rdfs_pyrophyllite_si}
\end{figure*}

Fig.~\ref{fig:rdfs_kaolinite_si} shows the radial distribution functions (RDFs) for all element pairs in kaolinite, using classical molecular dynamics and path integral molecular dynamics; Fig.~\ref{fig:rdfs_pyrophyllite_si} shows the same plots for pyrophyllite. In all cases, the effect of nuclear quantum fluctuations is quite modest, with the most pronounced effects seen for pairs involved in hydrogen bonding, as shown in the main text.

\clearpage

\section*{Hydrogen-Bond Population Correlation Functions}

\begin{figure}[hb]
    \centering
    \includegraphics[scale=0.7]{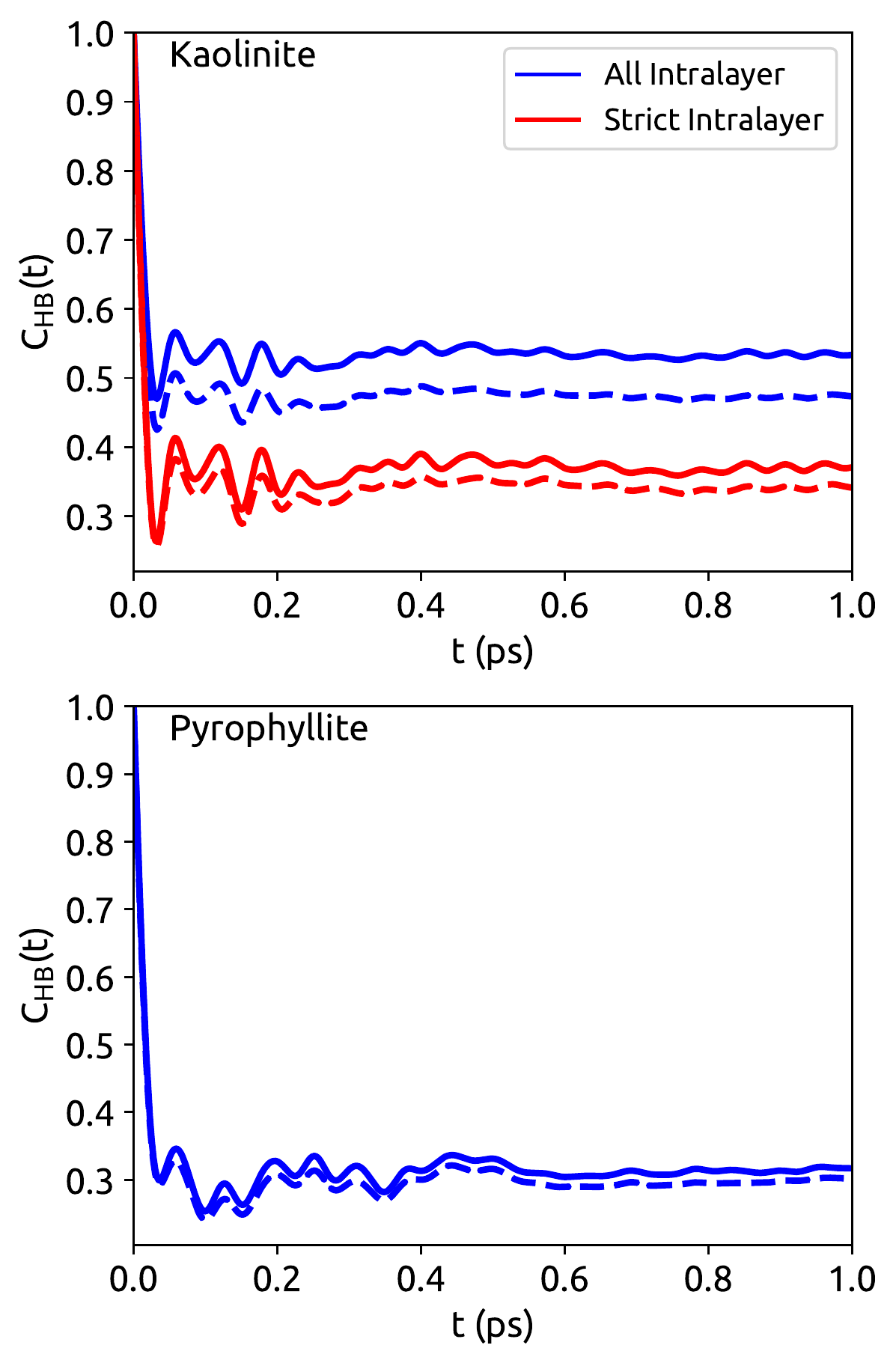}
    \caption{
    H-bond population correlation function for kaolinite (top panel) and pyrophyllite (bottom panel). Blue lines give this function for intralayer hydrogen bonds, and red lines for hydrogen bonds involving O--H bonds that never participate in interlayer hydrogen-bonding.
    Solid lines show the results of classical MD calculations and dashed lines the results of TRPMD.
    }
    \label{fig:hbond_populations_intralayer}
\end{figure}

Fig.~\ref{fig:hbond_populations_intralayer} shows the hydrogen bond population correlation function for kaolinite, both for all O--H bonds that are participating in intralayer H-bonds at $t=0$ and for O--H bonds that only participate in intralayer (i.e., no interlayer) H-bonds. These are compared to the H-bond population correlation function for pyrophyllite.

\noindent When the correlation function is calculated for O--H bonds in kaolinite that only participate in intralayer H-bonding (``Strict Intralayer''), the plateau value is much closer to that of the H-bonds in pyrophyllite, and the quantum effect is smaller than when all intralayer H-bonds are used to compute the function.
This indicates that a significant reason for the difference between intralayer H-bonds in kaolinite and those in pyrophyllite is simply that some of the former are able to to switch between intralayer and interlayer hydrogen bonds, with the latter incurring a larger quantum effect.

\clearpage

\section*{CLAYFF-TRPMD Properties}

The CLAYFF-TRPMD potential differs from the CLAYFF potential solely by the force constant for the O--H bond, which in the former is 2\% higher than in the latter. This leads to a significant change in the positions of the O--H stretching vibration in the vibrational spectra of clays. We show in this section that the effect on the other properties of kaolinite and pyrophyllite considered in the main text is negligible.

Table~\ref{tab:clay_trpmd_structure} shows the structural parameters for kaolinite and pyrophyllite from PIMD calculations, using the two models. The effect on the box lengths of modifying the forcefield (given by the parameter $\Delta$) is lower, often by an order of magnitude, than that of incorporating nuclear quantum effects into simulations with the original CLAYFF model (given by $\Delta_{\text{\tiny{QC}}}$), while the differences in unit cell angles are generally on the same order of magnitude.
Overall, the difference between the two forcefields is extremely small.
\begin{table}[h!]
\centering
\caption{
Unit cell parameters for kaolinite and pyrophyllite, as obtained from PIMD  simulations in the constant-stress ensemble, for the original CLAYFF model (ORIG) and the CLAYFF-TRPMD model (TRPMD).
Note that since a $2\times 2\times 2$ supercell is used, all lengths are divided by 2.
$\Delta$ is the difference between calculations with CLAYFF-TRPMD and the original CLAYFF model, and $\Delta_{\text{\tiny{QC}}}$ the difference between quantum and classical calculations for the original CLAYFF model (as reported in the main text).
In all cases, the uncertainty is in the fourth (for unit cell lengths) or third (for angles) decimal digit.
\label{tab:clay_trpmd_structure}}
\begin{tabular}{c|cccc|cccc}
\hline\hline
& \multicolumn{4}{c|}{Kaolinite} & \multicolumn{4}{c}{Pyrophyllite} \\
Property & ORIG & TRPMD & $\Delta$ & $\Delta_{\text{\tiny{QC}}}$ & ORIG & TRPMD & $\Delta$ & $\Delta_{\text{\tiny{QC}}}$ \\
\hline
$a (\text{\AA})$ & 5.204 & 5.205 & 0.001 & 0.009 & 5.202 & 5.201 & -0.001 & 0.011 \\
$b (\text{\AA})$ & 8.969 & 8.969 & 0.000 & 0.019 & 9.033 & 9.034 & 0.001 & 0.020 \\
$c (\text{\AA})$ & 7.444 & 7.445 & 0.001 & 0.018 & 9.430 & 9.422 & -0.080 & 0.018 \\
$\alpha (^{\circ})$ & 91.71 & 91.73 & 0.02 & 0.03 & 91.44 & 91.39 & -0.05 & 0.00 \\
$\beta (^{\circ})$ &  104.82 & 104.84 & 0.02 & -0.01 & 98.96 & 98.76 & -0.20 & 0.04 \\
$\gamma (^{\circ})$ & 90.56 & 90.57 & 0.01 & -0.03 & 89.82 & 89.82 & 0.00 & 0.00 \\
\hline\hline
\end{tabular}
\end{table}

\begin{figure*}
\centering
    \begin{subfigure}[b]{0.49\textwidth}
        \centering 
        \includegraphics[scale=0.7]{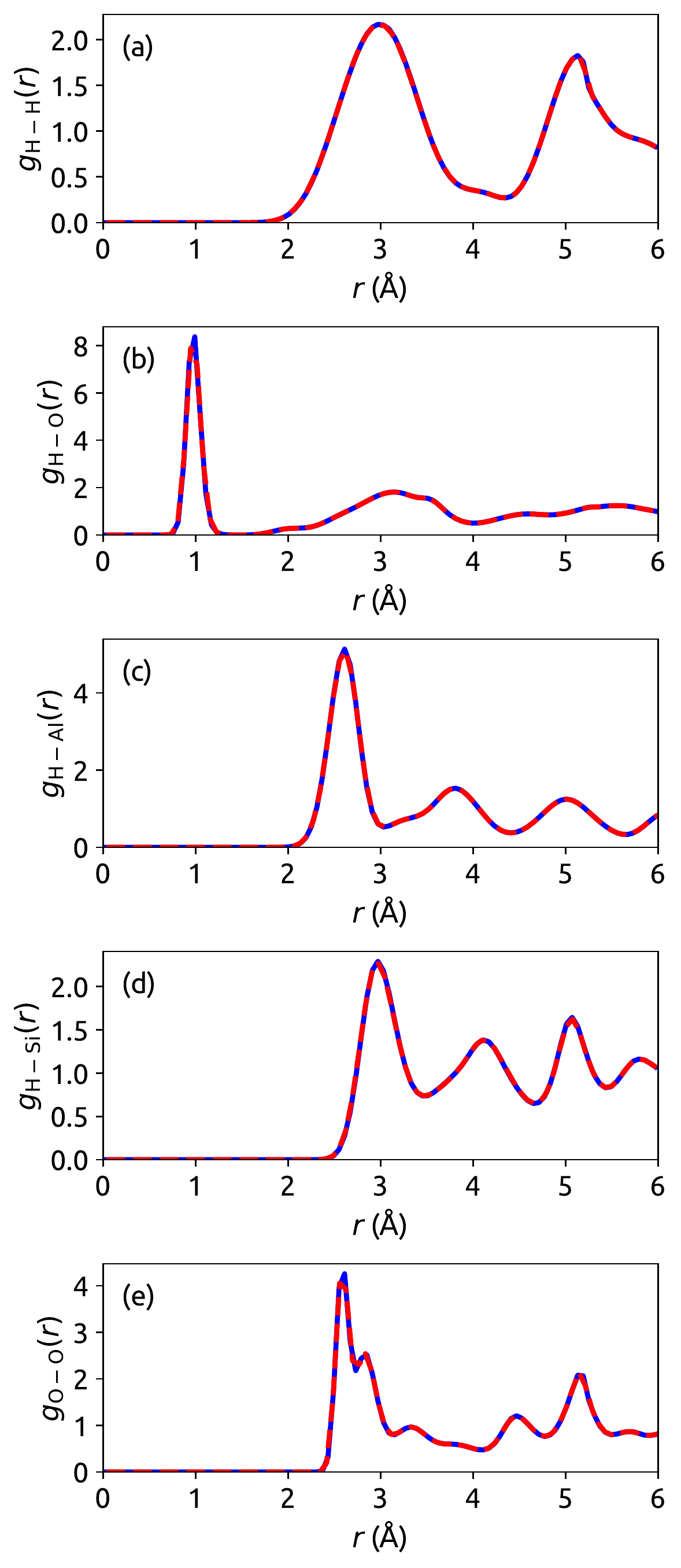}
        \label{subfig:rdfs_kaolinite_trpmd_1} 
    \end{subfigure}
    \begin{subfigure}[b]{0.49\textwidth}
        \centering 
        \includegraphics[scale=0.7]{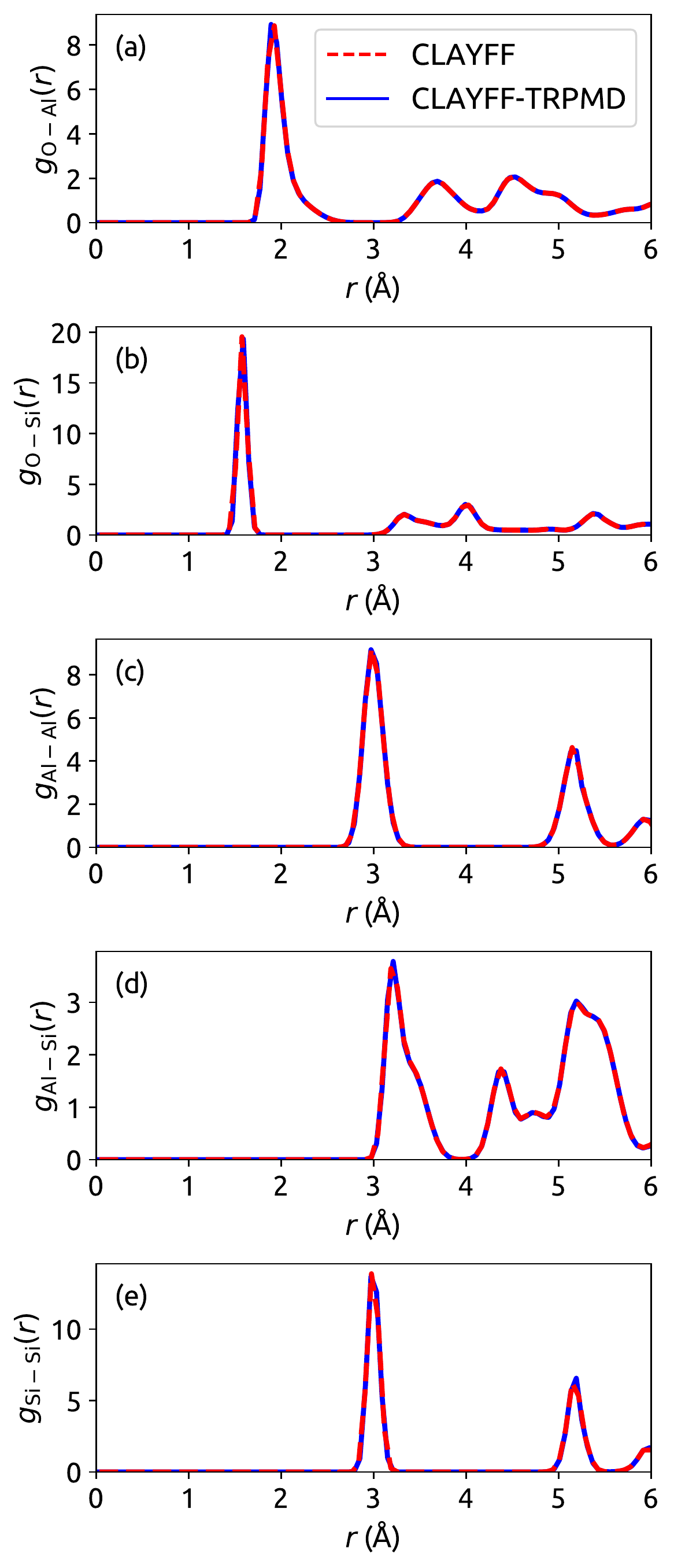}
        \label{subfig:rdfs_kaolinite_trpmd_2} 
    \end{subfigure}
    \caption{
    Radial distribution functions (RDFs) for all element pairs in kaolinite. Dashed red lines show the RDF when calculated using the original CLAYFF forcefield and solid blue lines show the RDF from the modified CLAYFF-TRPMD forcefield. In all cases, path integral molecular dynamics was used.
    } 
    \label{fig:rdfs_kaolinite_trpmd}
\end{figure*}

\begin{figure*}
\centering
    \begin{subfigure}[b]{0.49\textwidth}
        \centering 
        \includegraphics[scale=0.7]{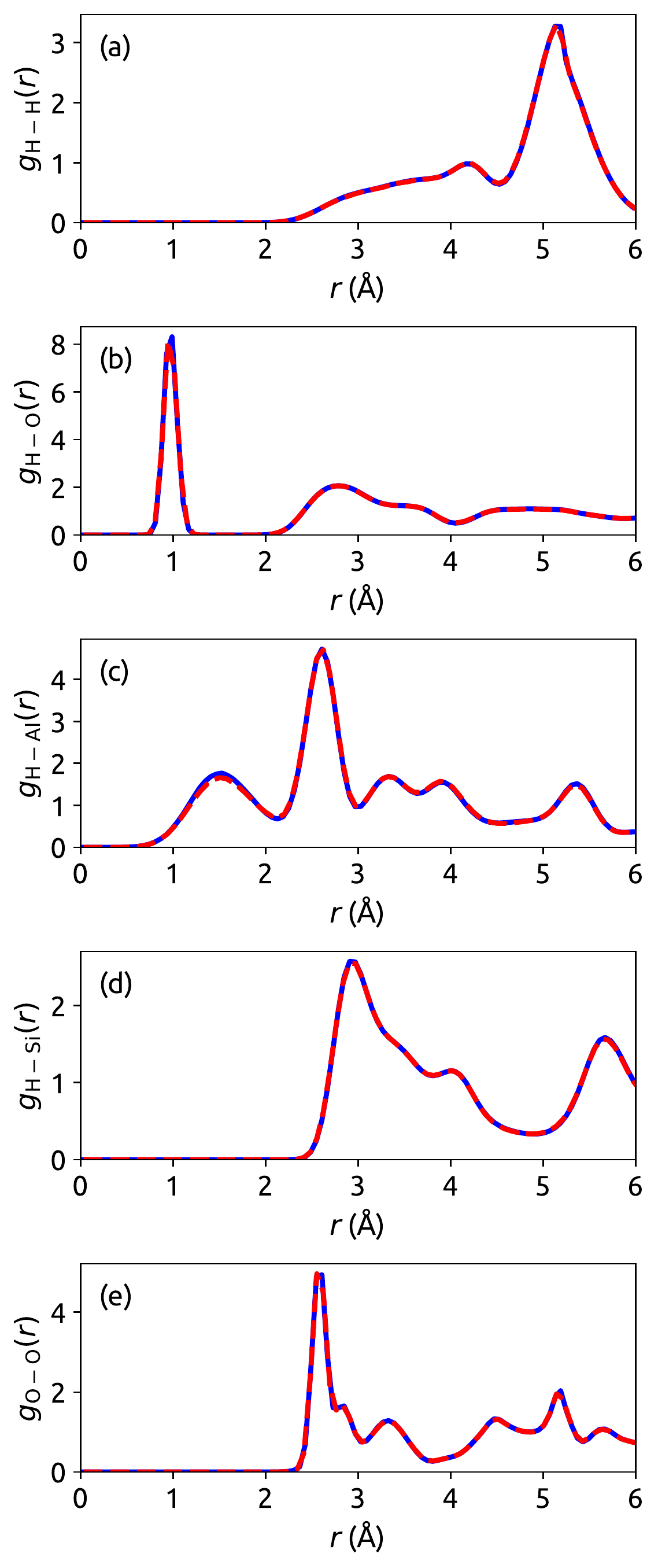}
        \label{subfig:rdfs_pyrophyllite_trpmd_1} 
    \end{subfigure}
    \begin{subfigure}[b]{0.49\textwidth}
        \centering 
        \includegraphics[scale=0.7]{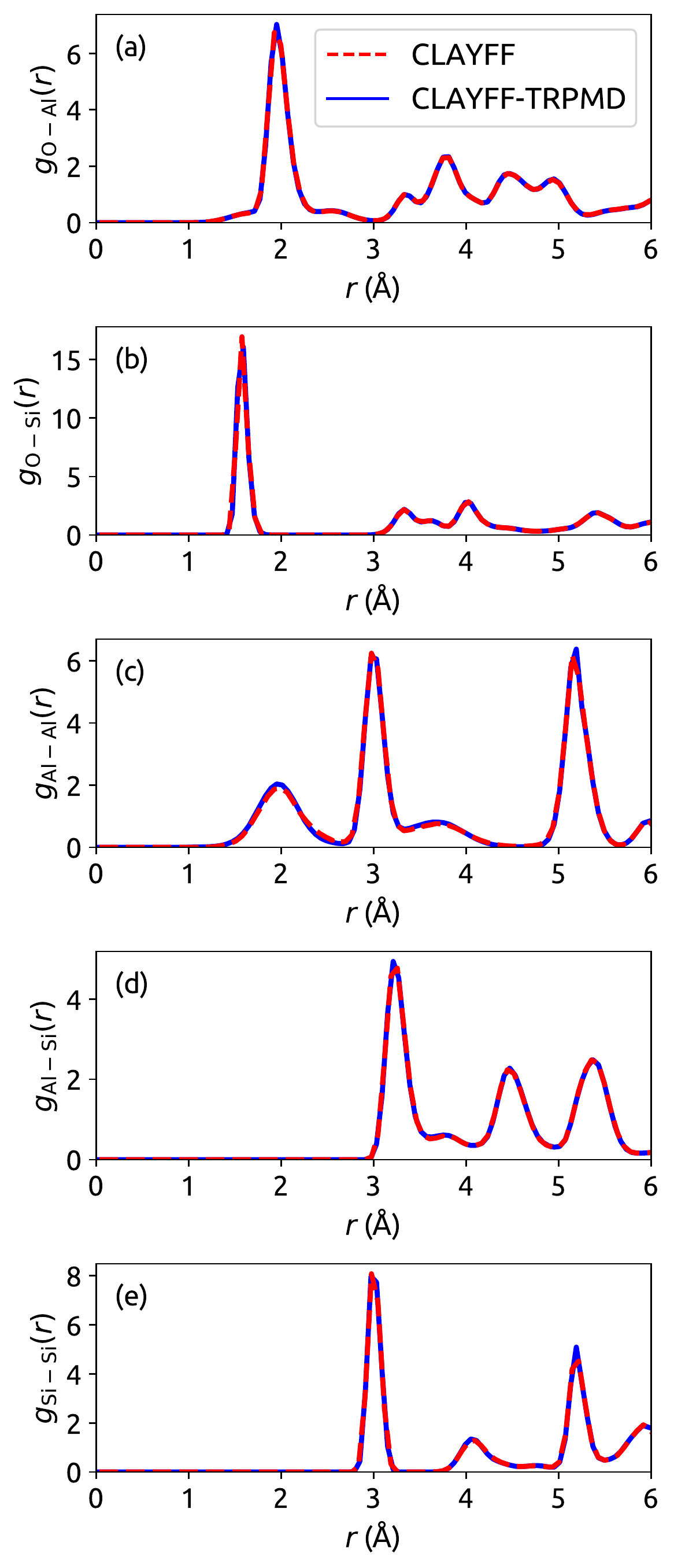}
        \label{subfig:rdfs_pyrophyllite_trpmd_2} 
    \end{subfigure}
    \caption{
    Radial distribution functions (RDFs) for all element pairs in pyrophyllite.
    Dashed red lines show the RDF when calculated using the original CLAYFF forcefield and solid blue lines show the RDF from the modified CLAYFF-TRPMD forcefield. In all cases, path integral molecular dynamics was used.
    } 
    \label{fig:rdfs_pyrophyllite_trpmd}
\end{figure*}

Fig.~\ref{fig:rdfs_kaolinite_trpmd} shows that the RDFs for kaolinite using CLAYFF and CLAYFF-TRPMD are essentially identical, and Fig.~\ref{fig:rdfs_pyrophyllite_trpmd} shows the same for pyrophyllite.
Table~\ref{tab:hbonds_trpmd} shows the average number of H-bonds formed in both models is also largely unchanged. Taken together with Table~\ref{tab:clay_trpmd_structure} and Figs.~\ref{fig:rdfs_kaolinite_trpmd} and \ref{fig:rdfs_pyrophyllite_trpmd}, these results show that the structural properties of the CLAYFF model are retained in the CLAYFF-TRPMD model.
\begin{table}[h!]
\centering
\caption{
Number of hydrogen bonds that are intact for kaolinite and pyrophyllite, from path integral MD calculations, using the original CLAYFF (ORIG) and CLAYFF-TRPMD (TRPMD) models. For kaolinite, H-bonds are classified as being intralayer or interlayer. ``$T=0~\text{K}$ Intralayer'' H-bonds are intralayer hydrogen bonds that exist at zero temperature, while ``$T>0~\text{K}$ Intralayer'' H-bonds are intralayer hydrogen bonds formed by O--H bonds that at zero temperature participate in \textit{interlayer} H-bonding.
For pyrophyllite, all H-bonds are intralayer.
\label{tab:hbonds_trpmd}}
\begin{tabular}{c|cc}
\hline\hline
\multicolumn{3}{c}{Kaolinite} \\
\hline\hline
Type & ORIG & TRPMD \\
Interlayer & 32.9 & 32.9 \\
$T=0~\text{K}$ Intralayer & 12.2 & 12.2 \\
$T>0~\text{K}$ Intralayer & 6.1 & 6.0 \\
\hline\hline
\multicolumn{3}{c}{Pyrophyllite} \\
\hline\hline
Type & ORIG & TRPMD \\
Intralayer & 28.6 & 28.5 \\
\hline\hline
\end{tabular}
\end{table}

\clearpage

Table~\ref{tab:kinetic_tensor_trpmd} shows the components of the quantum kinetic energy tensor for the CLAYFF and CLAYFF-TRPMD models.
Although the component of the quantum kinetic energy along the O--H bond is larger in the CLAYFF-TRPMD model, which has a stiffer bond and thus a larger zero-point energy, the differences in breaking hydrogen bonds are essentially unchanged and the conclusions of the main text still hold.
\begin{table}[htb!]
\centering
\caption{
Components of the centroid virial kinetic energy tensor $\boldsymbol{\mathcal{T}}$ for H atoms in intact hydrogen bonds or part of a dangling O--H bond, for intralayer and interlayer hydrogen-bonding in kaolinite and for intralayer hydrogen-bonding in pyrophyllite.
$\mathcal{T}_{\parallel}$ is the component parallel to the O--H bond, $\mathcal{T}_{\perp}$ the sum of the two components perpendicular to the bond, and $\mathcal{T}=\mathcal{T}_{\parallel} + \mathcal{T}_{\perp}$.
All values are in meV.
Results are given for the original CLAYFF forcefield and for the modified CLAYFF-TRPMD forcefield.
\label{tab:kinetic_tensor_trpmd}}
\begin{tabular}{c|ccc|ccc||ccc|ccc}
\hline\hline
\multicolumn{13}{c}{Kaolinite} \\
\hline\hline
& \multicolumn{6}{c||}{CLAYFF} & \multicolumn{6}{c}{CLAYFF-TRPMD} \\
\hline
& \multicolumn{3}{c|}{H-Bond} & \multicolumn{3}{c||}{Dangling Bond} & \multicolumn{3}{c|}{H-Bond} & \multicolumn{3}{c}{Dangling Bond} \\
\hline
Type & $T_{\parallel}$ & $\mathcal{T}_{\perp}$ & $\mathcal{T}$ & $\mathcal{T}_{\parallel}$ & $\mathcal{T}_{\perp}$ & $\mathcal{T}$ & $T_{\parallel}$ & $\mathcal{T}_{\perp}$ & $\mathcal{T}$ & $\mathcal{T}_{\parallel}$ & $\mathcal{T}_{\perp}$ & $\mathcal{T}$ \\
Intralayer & 100.7 & 40.6 & 141.3 & 100.4 & 40.9 & 141.3 & 102.6 & 40.7 & 143.3 & 102.4 & 41.0 & 143.4 \\
Interlayer & 100.7 & 40.3 & 141.0 & 102.2 & 35.8 & 138.0 & 102.7 & 40.3 & 143.0 & 104.3 & 35.9 & 140.2 \\
\hline\hline
\multicolumn{13}{c}{Pyrophyllite} \\
\hline\hline
& \multicolumn{6}{c||}{CLAYFF} & \multicolumn{6}{c}{CLAYFF-TRPMD} \\
\hline
& \multicolumn{3}{c|}{H-Bond} & \multicolumn{3}{c||}{Dangling Bond} & \multicolumn{3}{c|}{H-Bond} & \multicolumn{3}{c}{Dangling Bond} \\
\hline
Type & $\mathcal{T}_{\parallel}$ & $\mathcal{T}_{\perp}$ & $\mathcal{T}$ & $\mathcal{T}_{\parallel}$ & $\mathcal{T}_{\perp}$ & $\mathcal{T}$ & $T_{\parallel}$ & $\mathcal{T}_{\perp}$ & $\mathcal{T}$ & $\mathcal{T}_{\parallel}$ & $\mathcal{T}_{\perp}$ & $\mathcal{T}$ \\
Intralayer & 101.9 & 37.7 & 139.6 & 102.1 & 37.1 & 139.2 & 103.9 & 37.8 & 141.7 & 104.2 & 37.1 & 141.3 \\
\hline\hline
\end{tabular}
\end{table}

\begin{figure}[ht!]
\centering
\includegraphics[scale=0.7]{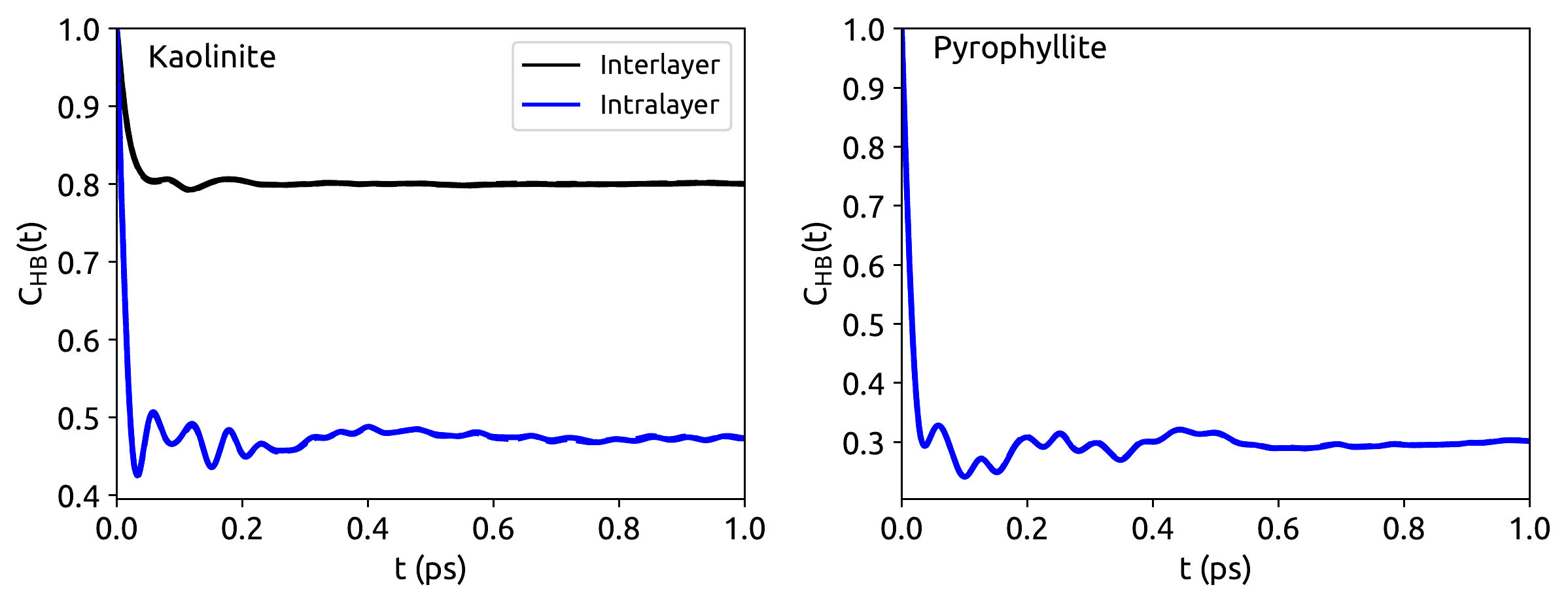}
\caption{
H-bond population correlation function of Eq.~(2) [main text], for kaolinite (left panel) and pyrophyllite (right panel). Black lines give this function for interlayer hydrogen bonds, where present, and blue lines for intralayer hydrogen bonds.
Solid lines show the results of TRPMD calculations using the CLAYFF model, and dashed lines the results of TRPMD calculations with the CLAYFF-TRPMD model.
\label{fig:hbond_correlation_function_trpd}
}
\end{figure}

Fig.~\ref{fig:hbond_correlation_function_trpd} shows that the hydrogen-bond population correlation function $C_{\text{\tiny{HB}}}(t)$ is indistinguishable when the CLAYFF model is replaced by the CLAYFF-TRPMD model.
Taken together, the results of this section show that the properties considered in the main text for CLAYFF are almost entirely unchanged when the force constant is increased, and that the only nonnegligible difference between the two models is the description of the vibrational spectrum.
